# Analysis of two-nucleon transfer reactions in the $^{20}$Ne + $^{116}$Cd system at 306 MeV

D. Carbone[1], J. L. Ferreira[2], S. Calabrese[1,3], F. Cappuzzello[1,3], M. Cavallaro[1], A. Hacisalihoglu[4], H. Lenske[5], J. Lubian[2], R. I. Magaña Vsevolodovna[6], E. Santopinto[6], C. Agodi[1], L. Acosta[7,8], D. Bonanno[8], T. Borello-Lewin[9], I. Boztosun[10], G. A. Brischetto[1,3], S. Burrello[1,11], D. Calvo[12], E.R. Chávez Lomelí[7], I. Ciraldo[1,3], M. Colonna[1], F. Delaunay[12,13,14], N. Deshmukh[15], P. Finocchiaro[1], M. Fisichella[1], A. Foti[8], G. Gallo[3,7], F. Iazzi[12,13], L. La Fauci[1,3], G. Lanzalone[1,16], R. Linares[2], N. H. Medina[9], M. Moralles[17], J.R.B. Oliveira[9], A. Pakou[18], L. Pandola[1], H. Petrascu[19], F. Pinna[12,13], S. Reito[8], G. Russo[3,8], O. Sgouros[1], S.O. Solakci[10], V. Soukeras[1], G. Souliotis[20], A. Spatafora[1,3], D. Torresi[1], S. Tudisco[1], A. Yildirin[10], V.A.B. Zagatto[2]

for the NUMEN collaboration

[1] INFN, Laboratori Nazionali del Sud, Catania, Italy

[2] Instituto de Fisica, Universidade Federal Fluminense, Niteroi, Brazil

[3] Dipartimento di Fisica e Astronomia, Università di Catania, Catania, Italy

[4] Institute of Natural Sciences, Karadeniz Teknik Universitesi, Trabzon, Turkey

[5] Department of Physics, University of Giessen, Germany

[6] INFN, Sezione di Genova, Genova, Italy

[7] Instituto de Fisica, Universidad Nacional Autónoma de México, Mexico City, Mexico

[8] INFN, Sezione di Catania, Catania, Italy

[9] Instituto de Fisica, Universidade de São Paulo, São Paulo, Brazil

[10] Departament of Physics, Akdeniz Universitesi, Antalya, Turkey

[11] Université Paris-Saclay, CNRS/IN2P3, IJCLab, 91405 Orsay, France

[12] INFN, Sezione di Torino, Torino, Italy

[13] DISAT, Politecnico di Torino, Torino, Italy

[14] LPC Caen, Normandie Université, ENSICAEN, UNICAEN, CNRS/INP3, France

[15] School of Sciences, Auro University, Surat, India

[16] Facoltà di Ingegneria e Architettura, Università di Enna "Kore", Enna, Italy

[17] Instituto de Pesquisas Energeticas e Nucleares IPEN/CNEN, São Paulo, Brazil

[18] Department of Physics, University of Ioannina and Hellenic Institute of Nuclear Physics, Ioannina, Greece

[19] IFIN-HH, Bucarest, Romania

[20] Department of Chemistry, University of Athens and Hellenic Institute of Nuclear Physics, Athens, Greece

**Abstract**

**Background:** Heavy-ion induced two-nucleon transfer reactions are powerful tools to reveal peculiar aspects of the atomic nucleus, such as pairing correlations, single-particle and collective degrees of freedom, and more. Also, these processes are in competition with the direct meson exchange in the double charge exchange reactions, which have recently attracted great interest due to their possible connection to neutrinoless double-beta decay. In this framework, the exploration of two-nucleon transfer reactions in the $^{20}$Ne + $^{116}$Cd collision at energies above the Coulomb barrier is particularly relevant since the $^{116}$Cd nucleus is a candidate for the double-β decay.

**Purpose:** We want to analyse selected transitions to low-lying $0^+$ and $2^+$ states of the residual nuclei in the $^{116}$Cd($^{20}$Ne,$^{22}$Ne)$^{114}$Cd two-neutron pickup and $^{116}$Cd($^{20}$Ne,$^{18}$O)$^{118}$Sn two-proton stripping reactions at 306 MeV incident energy and determine the role of the couplings with inelastic transitions.

**Methods:** We measured the excitation energy spectra and absolute cross sections for the two reactions using the MAGNEX large acceptance magnetic spectrometer to detect the ejectiles. We performed direct coupled reaction channels and sequential distorted wave Born approximation calculations using the double folding São Paulo potential to model the initial and final state interactions. The spectroscopic amplitudes for two- and single-particle transitions were derived by different nuclear structure approaches: microscopic large-scale shell model, interacting boson model-2 and quasiparticle random phase approximation.

**Results**: The calculations are able to reproduce the experimental cross sections for both two-neutron and two-proton transfer reactions provided that the couplings with the inelastic channels are taken into account. A competition between the direct and the sequential process is found in the reaction mechanism. For the two-proton transfer case, the inclusion of the $1g_{7/2}$ and $2d_{5/2}$ orbitals in the model space is crucial.

## 1. Introduction

Heavy-ion multi-nucleon transfer reactions have been extensively studied during the last years [1] [2] [3] [4] [5] [6] [7] [8], revealing interesting phenomena connected to single-particle, pairing correlations and cluster degrees of freedom. Moreover, these studies are also complementary to those on Double Charge Exchange (DCE) reactions, which have recently attracted interest for their possible connection to neutrinoless double-beta decay [9] [10] [11] [12] [13]. In particular, from the multi-nucleon transfer studies, it is possible to obtain important information on the nuclear wave functions and on the role of the mean-field dynamics in DCE reactions [14]. The latter is a competitive mechanism to the meson exchange involved in DCE reactions [15] [16].

A complete treatment of the transfer process, which contains the one-step channel with the inclusion of all the possible inelastic excitations of the involved nuclei, the sequential channel, with the inclusion of intermediate partitions, and the non-orthogonal term, is still not available in the state-of-the-art theories. A possible way to describe the two-neutron transfer reactions is the second order Distorted Wave Born Approximation (DWBA) approach, recently applied in refs. [17] [18] for (p,t) reactions on tin isotopes. Nevertheless, this approach does not include the inelastic excitations, relevant when dealing with heavy-ion induced reactions [19] [20] [21]. A possible way to treat the reaction mechanism is to include explicitly the inelastic excitations by using the coupled-channels approach for the one-step channel, including the corresponding non-orthogonal term, and perform separately the two-step calculations. The results can be then summed by considering the relative phase as an additional parameter. Within this approach, interesting results were found for the $^{18}$O induced one- and two-neutron transfer reactions. For a long time, the approximations used to deal with the complex many-body aspects of the reactions led to the use of arbitrary scaling factors in the calculated cross sections in order to compare them with the experimental results [22] [23], preventing the extraction of accurate nuclear structure information. With the advent of microscopic approaches based on DWBA and Coupled Channels (CC) schemes with double-folding potentials and spectroscopic amplitudes derived from Large Scale Shell Model (SM) or Interacting Boson Model (IBM) [24], it was possible to give a satisfactory description of the measured cross sections [25] [26] [27] [28] [29] [30] [31].

In this context, another relevant approach is the Quasiparticle Random Phase Approximation (QRPA), which gives a realistic description of collectivity in nuclear response functions. In modern QRPA calculations extremely large two-quasiparticle (2QP), i.e. one-particle-one-hole, configuration spaces are covered by which the (energy weighted) sum rules of transition operators are fully exhausted. In this sense, QRPA methods are more suitable with respect to shell model calculations which notoriously underestimate collective enhancements observed in nuclear strength functions. These enhancements by phase-coherence will play also a major role in transfer reactions populating the daughter nucleus excited states of large collectivity, e.g. the lowest $2^+$-states. Such investigations, both experimentally and theoretically, are a new approach to reveal the collectivity of low-lying nuclear states as an important supplementary aspect to the primary interest on gaining insight into nuclear pair dynamics. In the past, QRPA methods have been extensively used for single-particle transfer and charge exchange reactions induced by light and heavy ions, e.g. [32] [33] [34]. A detailed overview is found in the recent review article [16].

Presently, such newly developed techniques are general enough to be extended to other heavy-ion induced transfer reactions for which much less is known. In this context, much interest is raising for studying reactions induced by $^{20}$Ne beams [35] [36], for which, to our knowledge, no experimental data existed at beam energies and mass region of interest for the NUMEN project [13].

In this paper, we show a consistent study of two-neutron and two-proton transfer induced in $^{20}$Ne + $^{116}$Cd collisions at 15 MeV/u incident energy, investigated under the same experimental conditions and the same theoretical framework applied for $^{18}$O induced reactions [27] [30]. We analyse new data concerning the low-lying states populated in the $^{116}$Cd($^{20}$Ne,$^{18}$O)$^{118}$Sn and $^{116}$Cd($^{20}$Ne,$^{22}$Ne)$^{114}$Cd reactions within DWBA and Coupled Reaction Channel (CRC) using SM, Interacting Boson Model-2 (IBM-2) and QRPA spectroscopic amplitudes. To our knowledge, this is the first time that such a broad theoretical framework is applied to the two-proton transfer channel and that the $^{20}$Ne induced transfer reactions are explored. Such a description is complementary to the earlier shell model and pairing methods, see e.g. [3] [37].

## 2. Experimental setup and results

The experiment has been performed at the INFN-LNS laboratory in Catania in the framework of the NUMEN project [13]. The $^{20}$Ne$^{4+}$ beam, accelerated at 306 MeV incident energy by the K800 Superconducting Cyclotron, was fully stripped by crossing a thin carbon foil and transported to impinge on a 1370 ± 70 μg/cm$^2$ $^{116}$Cd target in the case of the two-proton transfer and a 1080 ± 60 μg/cm$^2$ $^{116}$Cd target in the case of two-neutron transfer. Both foils are 96% isotopically enriched and produced by rolling at the LNS thin film laboratory. A total charge of 430 ± 40 μC for the two-proton case and 530 ± 50 μC in the two-neutron one was collected by a Faraday cup mounted 15 cm downstream of the target. The ejectiles were momentum analysed by the MAGNEX spectrometer [38] [39] [40] in separated runs. For the $^{116}$Cd($^{20}$Ne,$^{18}$O)$^{118}$Sn reaction the optical axis of the spectrometer was placed at $\theta_{lab}$ = 8° in the laboratory frame with an angular acceptance of $\Omega$ ~ 45 msr. For the $^{116}$Cd($^{20}$Ne,$^{22}$Ne)$^{114}$Cd reaction MAGNEX was placed at $\theta_{lab}$ = 9° and the vertical acceptance was reduced, with a corresponding total solid angle of Ω ~ 1.3 msr, in order to reduce the large overall detection rate at the focal plane detector. The measured angular range is 3° < $\theta_{lab}$ < 14° and 4° < $\theta_{lab}$ < 15°, respectively. The magnetic fields of the magnetic elements were set in order to focus the ejectiles corresponding to the population of the $^{118}$Sn$_{g.s.}$ in the two-proton transfer case at $\delta(^{118}\mathrm{Sn}_{g.s.}) = (p - p_0) / p_0 = -0.043$ (where $\delta$ represents the fractional deviation of the momentum $p$ from the reference one $p_0$). Given the momentum acceptance of the spectrometer, an excitation energy spectrum of $^{118}$Sn up to $E_x$ ~ 16 MeV was explored. In the $^{116}$Cd($^{20}$Ne,$^{22}$Ne)$^{114}$Cd case, $\delta(^{114}Cd_{g.s.}) = -0.035$ was set, which corresponds to a maximum excitation energy of $^{114}$Cd $E_x \sim 24$ MeV.

The ejectile identification and the data reduction techniques are the same described in details in refs. [41] [42] [43]. The latter is based on a fully differential algebraic method [44] and requires the measured horizontal and vertical positions and angles at the focal plane as input [45]. Examples of the obtained energy spectra for the $^{118}$Sn and $^{114}$Cd residual nuclei are shown in Figure 1 in which $E_x = Q_0 - Q$, where $Q_0$ is the ground-state-to-ground-state reaction $Q$-value. An energy resolution of ~ 300 keV full width at half maximum is obtained for the two-neutron spectrum, whereas it is ~ 800 keV for the two-proton case. The worse energy resolution obtained in the two-proton transfer case comes from the target thickness effect, which is dominant when the atomic number $Z$ is changed between the beam and the ejectile.

The absolute cross sections were extracted according to the technique described in ref. [42], taking into account the overall MAGNEX efficiency [46]. The error bars included in the spectra indicate the statistical uncertainty. An overall uncertainty of ~ 10%, not shown in Figure 1, is common to all the points in the spectra, originating from the target thickness measurement and the Faraday cup charge collection.

The continuum shape of the two spectra shown in Figure 1 is a composition of the high-level density of the involved heavy nuclei and the limited energy resolution. The energy distributions show pronounced maxima at comparatively high excitation energy ($E_x \sim 15$ MeV for two-proton and $E_x \sim 13$ MeV for two-neutron).

In the zoomed view of the two spectra, shown in the insets of Figure 1, the ground and the first excited states of the ejectile and residual nuclei are visible, populated with very low yields due to unfavourable matching conditions for these low energy and low angular momentum transitions [47]. The integrated values of the measured cross sections for the transitions to the ground and low-lying excited states of the residual nucleus were taken as: (i) the area of each Gaussian function fit, shown in the inset of Figure 1(upper panel), for the two-proton transfer reaction; (ii) integration of the counts in the $-0.2 \lesssim E_x \lesssim 0.2$ MeV region for the g.s. and $0.4 \lesssim E_x \lesssim 0.8$ MeV for the first excited state of $^{114}$Cd at $E_x = 0.558$ MeV for the two-neutron transfer case. The results are listed in Table 1. The uncertainty is dominated by the statistical contribution.

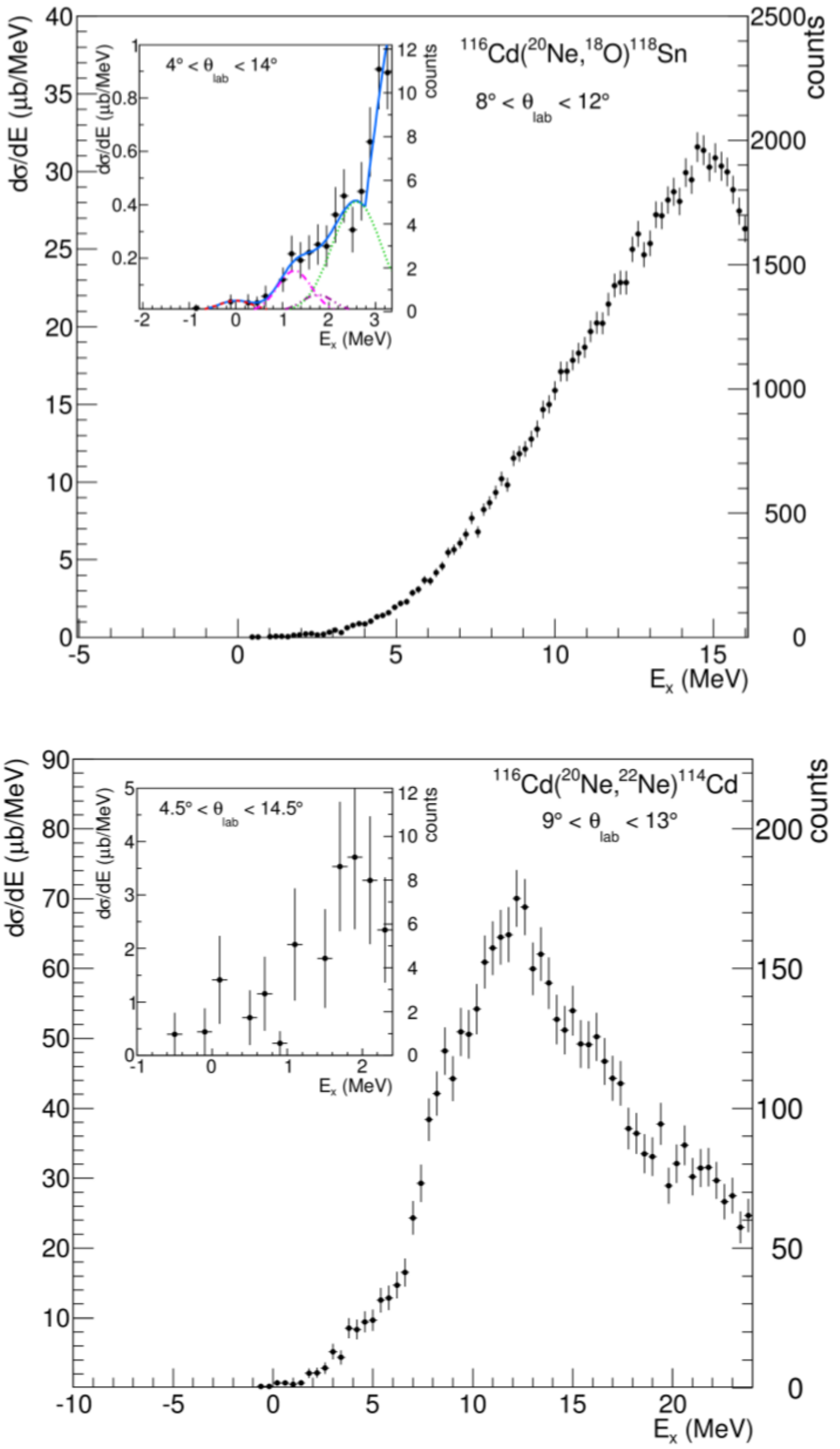

**Figure 1**. Upper panel: excitation energy spectra of the $^{116}$Cd($^{20}$Ne,$^{18}$O)$^{118}$Sn two-proton transfer reaction at 306 MeV and 8° < $\theta_{lab}$ < 12°. Inset: zoomed view of the low-lying states for 4° < $\theta_{lab}$ < 14°. Lines obtained from best-fit procedures identify transitions to particular states: ground state ($0^+$) (red dashed-line), 1.229 MeV ($2^+$) (magenta dashed-dotted line), 1.758 MeV ($0^+$) (violet dashed-double-dotted line), a mixture of states between 2 and 3 MeV (green dotted-line) and the global result (blue line) that includes a background curve for the high level density above ~ 3 MeV. Lower panel: excitation energy spectra of the $^{116}$Cd($^{20}$Ne,$^{22}$Ne)$^{114}$Cd two-neutron transfer reaction at 306 MeV and 9° < $\theta_{lab}$ < 13°. Inset: zoomed view of the low-lying states for 4.5° < $\theta_{lab}$ < 14.5°.

**Table 1**. Integrated experimental cross sections for 4° < $\theta_{lab}$ < 14° in the two-proton transfer and 4.5° < $\theta_{lab}$ < 14.5° in the two-neutron transfer reactions.

| Final system | Exp. cross section (nb) |
|---|---|
| $^{18}O_{gs}(0^+)$ + $^{118}Sn_{gs}(0^+)$ | 40 ± 15 |
| $^{18}O_{gs}(0^+)$ + $^{118}Sn_{1.230}(2^+)$ | 140 ± 60 |
| $^{18}O_{gs}(0^+)$ + $^{118}Sn_{1.758}(0^+)$ | 60 ± 40 |
| $^{18}O_{gs}(0^+)$ + $^{118}Sn_{2\text{-}3\ MeV}$ | 540 ± 320 |
| $^{22}Ne_{gs}(0^+)$ + $^{114}Cd_{gs}(0^+)$ | 370 ± 190 |
| $^{22}Ne_{gs}(0^+)$ + $^{114}Cd_{0.558}(2^+)$ | 420 ± 170 |

## 3. Theoretical analysis

We performed microscopic calculations for the $^{116}$Cd($^{20}$Ne,$^{18}$O)$^{118}$Sn and $^{116}$Cd($^{20}$Ne,$^{22}$Ne)$^{114}$Cd reactions at 306 MeV incident energy to describe the transfer cross sections for the transitions to some of the experimentally populated low-lying states. The calculations were carried out considering the double folding São Paulo Potential as the optical potential for the initial and final state nucleus-nucleus interaction.

Studies on the São Paulo double folding systematics [48] have provided the real and imaginary normalization coefficients $N_r$ = 1.0 and $N_I$ = 0.78 [U(R) ≈ ($N_r$ + i$N_I$)$V_{LE}^{SP}$(R)], respectively, in describing the elastic scattering angular distributions for many systems in a wide mass and energy ranges (outside the region of strong couplings among different reaction channels) [49] [50]. The normalization coefficient of the imaginary part effectively considers the coupling of all the other channels to the elastic one, corresponding to the dissipative processes which absorb flux from the elastic scattering. On the other hand, when relevant inelastic states are explicitly coupled to the ground states of the projectile and/or target, this factor is further reduced to account for others possible channels (like continuum states and the fusion) that were not explicitly included in the coupled channel-coupling scheme.

The prior form of the potential was used to calculate the matrix elements, and the non-orthogonality corrections were introduced in the calculations. The single-particle bound states were generated by Woods-Saxon potentials, assuming $r$ = 1.26 fm and $a$ = 0.70 fm for the lighter nuclei cores and $r$ = 1.20 fm and $a$ = 0.60 fm for the heavier ones. The depth of these potentials was varied to fit the experimental binding energies of each valence nucleon. All the theoretical cross sections were calculated using the FRESCO code [51].

As our objective is to show microscopic results for the two-neutron pickup and two-proton stripping transfer cross-sections, the independent coordinates scheme and sequential methods were considered. The former assumes that the two nucleons are directly (simultaneously) transferred from the initial partition to the final one. A coordinate transformation is performed from the individual coordinate of each valence nucleon to the coordinate of their centre of mass relative to the core and the coordinate corresponding to their relative motion. These new coordinates account properly for the intrinsic two-particle states of the two transferred nucleons. The sequential method considers the two nucleons being transferred one by one passing through an intermediate partition.

The reaction calculations are connected to the structure of the involved nuclear states by the corresponding single- and two-particle spectroscopic amplitudes, respectively. They are derived microscopically by large-scale SM, IBM-2 and QRPA calculations.

### 3.1 Shell Model calculations

To obtain the spectroscopic amplitudes for both projectiles and target overlaps within the shell-model framework, the NushellX [52] code was used. A constrained model sub-space is frequently adopted to perform this kind of structure calculations because of the complexity in diagonalizing the Hamiltonian of systems involving open shell medium and heavy nuclei.

For the projectile overlaps, the *p-sd-mod* [53] phenomenological interaction was considered in the shell-model calculations. This interaction takes into account the full *p-sd* valence sub-space for protons and neutrons with $^{4}$He as a closed core and the valence orbits: $1p_{3/2}$, $1p_{1/2}$, $1d_{5/2}$, $2s_{1/2}$ and $1d_{3/2}$. The two-body matrix elements of the *p-sd-mod* interaction are a modified version of the ones introduced by Warburton and Brown in the PSDWBT interaction [54] for the *p-sd* model.

The model space used to describe the structural characteristics of the target and residual nuclei considers the $1f_{5/2}$, $2p_{3/2}$, $2p_{1/2}$ and $1g_{9/2}$ orbits for the valence protons, and the $1g_{7/2}$, $2d_{5/2}$, $2d_{3/2}$, $3s_{1/2}$ and $1h_{11/2}$ orbits for the valence neutrons. The effective interaction derived in this model space (named by *jj45pna* interaction) was elaborated using the $^{78}$Ni nucleus as a core. The proton-proton, neutron-neutron, and proton-neutron interactions were derived from the charge-dependent Bonn potential (CD-Bonn) based on the predictions of the Bonn Full Model [55] used in the description of the nucleon-nucleon interaction. It has been used to investigate the charge-symmetry-breaking and charge-independence-breaking effects corresponding to the meson-exchange processes [56]. In the present structure calculations, the orbit $1h_{11/2}$ was not considered in the transfer calculation because the model sub-space results too large and difficult to handle.

In order to verify the relevance of the $1g_{7/2}$ and $2d_{5/2}$ orbitals, above the full shell $fp$-$g_{9/2}$, on the two-proton transfer cross sections, a different valence model sub-space considering the $2p_{1/2}$, $1g_{9/2}$, $1g_{7/2}$ and $2d_{5/2}$ orbits for protons was used. The neutron sub-space is the same as the one of the *jj45pna* interaction. In this model space, the two-body matrix elements were obtained considering the CD-Bonn nucleon-nucleon potential from the effective shell-model Hamiltonian. The $^{88}$Sr nucleus is considered as a core [57], and because of this, we will call it as *88Sr45* interaction.

In Table 2 and 3, the comparison between the theoretical and experimental excitation energies of the low-lying states for all the involved nuclei in both $^{116}$Cd($^{20}$Ne,$^{22}$Ne)$^{114}$Cd and $^{116}$Cd($^{20}$Ne,$^{18}$O)$^{118}$Sn transfer reactions is shown. One can see a reasonably good agreement between theoretical and experimental spectra for both light and heavy nuclei.

**Table 2.** Comparison between the theoretical and experimental low-lying spectra obtained by shell model calculations for projectiles and ejectiles involved in the studied reactions. Energies are in MeV.

| Shell model: psdmod interaction | | | | | | | | | | | | | | |
|---|---|---|---|---|---|---|---|---|---|---|---|---|---|---|
| $^{20}$Ne | Exp. | Th. | $^{21}$Ne | Exp. | Th. | $^{22}$Ne | Exp | Th. | $^{19}$F | Exp. | Th. | $^{18}$O | Exp. | Th. |
| 0+ | 0 | 0 | 3/2+ | 0 | 0.238 | 0+ | 0 | 0 | 1/2+ | 0 | 0.108 | 0+ | 0 | 0 |
| 2+ | 1.634 | 2.253 | 5/2+ | 0.351 | 0.0 | 2+ | 1.275 | 1.731 | 1/2- | 0.110 | 0.756 | 2+ | 1.982 | 2.264 |
| 4+ | 4.248 | 4.594 | 7/2+ | 1.750 | 1.821 | 4+ | 3.357 | 3.560 | 5/2+ | 0.197 | 0.0 | 4+ | 3.555 | 3.621 |
| 2- | 4.967 | 4.990 | 1/2- | 2.789 | 2.230 | 2+ | 4.456 | 4.210 | 5/2- | 1.346 | 2.432 | 0+ | 3.634 | 4.251 |
| 3- | 5.621 | 5.328 | 1/2+ | 2.794 | 1.880 | 2- | 5.146 | 5.295 | 3/2- | 1.459 | 2.624 | 2+ | 3.920 | 4.188 |
| 1- | 5.788 | 8.736 | 9/2+ | 2.867 | 2.802 | 1+ | 5.330 | 5.430 | 3/2+ | 1.554 | 1.081 | 1- | 4.456 | 4.959 |

**Table 3.** Comparison between the theoretical and experimental low-lying spectra obtained by shell model calculations for the target and residual nuclei involved in the studied reactions. Energies are in MeV.

| Shell model: jj45pna interaction | | | | | | | | |
|---|---|---|---|---|---|---|---|---|
| $^{116}$Cd | Exp. | Th. | $^{115}$Cd | Exp. | Th. | $^{114}$Cd | Exp | Th. |
| 0+ | 0 | 0 | 1/2+ | 0 | 0.325 | 0+ | 0 | 0 |
| 2+ | 0.513 | 0.740 | (11/2)- | 0.181 | 2.195 | 2+ | 0.558 | 0.604 |
| 2+ | 1.213 | 1.782 | (3/2)+ | 0.229 | 0.0 | 0+ | 1.135 | 1.264 |
| 4+ | 1.219 | 1.712 | (5/2)+ | 0.361 | 0.534 | 2+ | 1.210 | 1.074 |
| 0+ | 1.283 | 1.526 | (7/2)- | 0.394 | 1.879 | 4+ | 1.284 | 1.543 |
| 0+ | 1.380 | 2.949 | (9/2)- | 0.417 | 2.141 | 0+ | 1.305 | 2.117 |
| Shell model: 88Sr45 interaction | | | | | | | | |
| $^{116}$Cd | Exp. | Th. | $^{117}$In | Exp | Th. | $^{118}$Sn | Exp. | Th. |
| 0+ | 0 | 0 | 9/2+ | 0 | 0 | 0+ | 0 | 0 |
| 2+ | 0.513 | 0.721 | 1/2- | 0.315 | 0.078 | 2+ | 1.230 | 0.802 |
| 2+ | 1.213 | 1.284 | 3/2- | 0.589 | 1.023 | 0+ | 1.758 | 3.474 |
| 4+ | 1.219 | 1.653 | 3/2+ | 0.660 | 2.147 | 2+ | 2.043 | 2.018 |
| 0+ | 1.283 | 2.032 | 7/2+ | 0.748 | 1.082 | 0+ | 2.057 | 4.304 |
| 0+ | 1.380 | 2.745 | 1/2+ | 0.749 | 2.011 | 4+ | 2.280 | 2.936 |

### 3.2 Interacting Boson Model 2

The microscopic IBM-2 is a way to calculate matrix elements for medium and heavy nuclei, that has been applied recently in neutrinoless double beta decay [58], nuclear matrix elements for double charge exchange [59], and two neutron transfer [27]. The nuclear wave functions are generated by diagonalizing the IBM-2 Hamiltonian [60]. The parameters of the even-even nuclei $^{114,116}$Cd and

$^{118}$Sn are taken from ref. [61]. The low-lying states of those nuclei are in quite good agreement with the experimental data, as it is shown in Table 4.

Two-nucleon transfer is modelled as a combination of two-neutron (proton) stripping and two-proton (neutron) pickup reactions [27]. The previous process can be described in terms of two-nucleon transfer operator. The target two-nucleon transfer spectroscopic amplitudes are calculated by using microscopic IBM-2 which corresponds to assuming that the matrix elements between fermionic states - in the collective subspace - are identical to the matrix elements in the bosonic space, so the matrix elements of the two nucleon transfer operators, in the Generalized Seniority scheme [62], are mapped into matrix elements of bosonic operators by the Otsuka, Arima and Iachello (OAI) method [63]. The mapping coefficients depend on structure coefficients that can be estimated by diagonalizing a surface delta pairing interaction [64] in the appropriate shell model sub-space, as in refs. [58] [59] [27].

The spectroscopic amplitudes were computed for the $\langle ^{114}\mathrm{Cd}|^{116}\mathrm{Cd}\rangle$ and $\langle ^{118}\mathrm{Sn}|^{116}\mathrm{Cd}\rangle$ overlaps considering a larger space than in the case of SM calculations, since it includes the $1f_{5/2}$, $2p_{3/2}$, $2p_{1/2}$, $1g_{9/2}$, $1g_{7/2}$, $2d_{5/2}$ orbitals as valence sub-space for the protons and the $1g_{7/2}$, $2d_{5/2}$, $2d_{3/2}$, $3s_{1/2}$ and $1h_{11/2}$ orbitals for the neutrons.

**Table 4.** Comparison between calculated and experimental low-lying states for the $^{116}$Cd, $^{114}$Cd and $^{118}$Sn nuclei. Energies are in MeV.

| Interacting Boson Model-2 | | | | | | | | |
|---|---|---|---|---|---|---|---|---|
| **$^{116}$Cd** | **Exp.** | **Th.** | **$^{114}$Cd** | **Exp** | **Th.** | **$^{118}$Sn** | **Exp.** | **Th.** |
| 0+ | 0 | 0 | 0+ | 0 | 0 | 0+ | 0 | 0 |
| 2+ | 0.513 | 0.516 | 2+ | 0.558 | 0.492 | 2+ | 1.230 | 1.201 |
| 2+ | 1.213 | 1.178 | 0+ | 1.135 | 1.274 | 0+ | 1.758 | 1.790 |
| 4+ | 1.219 | 1.186 | 2+ | 1.210 | 1.125 | 2+ | 2.043 | 2.261 |
| 0+ | 1.283 | 1.325 | 4+ | 1.284 | 1.130 | 4+ | 2.280 | 2.267 |

### 3.3 Quasi-Particle Random Phase Approximation

For the two-proton transfer case, we derived the spectroscopic amplitudes also using the QRPA approach in the target/residual nuclei. A self-consistent approach is used by describing nuclear ground states in Hartree-Fock-Bogoliubov theory and excited states by QRPA theory as coherent superpositions of 2QP-excitations. Throughout, we assume spherical symmetry. Interactions derived from Brueckner G-matrix calculations are used, supplemented by additional density-dependent three-body terms, as discussed in [65] [66] [67]. In Table 5 results for binding energies, excitation energies, and B(E2)-values are displayed and compared to data.

**Table 5**: Comparison of HFB and QRPA results for $^{116}$Cd and $^{118}$Sn to data. In the second column, HFB and measured binding energies, taken from the AMDE-2012 compilation [68], are displayed. The observed excitation energies and B(E2)-values are from [69].

| Nucleus | B(A)/A [MeV/nucleon] | $E_{th}(2^+)$ [MeV] | $B_{th}$(E2) [e2b2] | $E_{exp}(2^+)$ [MeV] | $B_{exp}$(E2) [$e^2b^2$] |
|---|---|---|---|---|---|
| $^{116}$Cd | 8.483 (th.) / 8.512 (exp.) | 0.520 | 0.564 | 0.513 | 0.501…0.680 |
| $^{114}$Cd | 8.468 (th.) / 8.488 (exp.) | 0.483 | 0.590 | 0.488 | 0.578(44) |
| $^{118}$Sn | 8.493 (th.) / 8.523 (exp.) | 1.231 | 0.211 | 1.2296 | 0.156…0.240 |

Here, we briefly sketch the derivation of two-particle transfer spectroscopic amplitudes for reactions populating nuclear states dominated by QRPA 2QP-configurations. The main purpose of the discussion is to show that the collective, phase-coherent features of QRPA transitions enter into the spectroscopic amplitudes of two-particle transfer reactions. In a conveniently chosen single particle representation, the pair addition multipole operators are given by the two-particle field operators

$$\Psi_{JM}(r_1, r_2) = \sum_{ik}\left(\phi_{JM}^{(j_i j_k)}(r_1, r_2) P_{JM}^{+}(j_i j_k) + \tilde{\phi}_{JM}^{(j_i j_k)}(r_1, r_2)\tilde{P}_{JM}(j_i j_k)\right), \quad (1)$$

with the two-particle creation operators $P_{JM}^{+}(j_i j_k) = \left[a_{j_i}^{+} a_{j_k}^{+}\right]_{JM}$ where angular momentum coupling is indicated by the bracket-notation and $\tilde{P}_{JM}(j_i j_k) = (-)^{J+M} A_{J-M}(j_i j_k)$ is the time-reversed (annihilation) operator, describing two-particle removal processes. We have introduced the two-particle multipole wave functions

$$\phi_{JM}^{(j_i j_k)}(r_1, r_2) = \left[\psi_{j_i}(r_1)\psi_{j_k}(r_2)\right]_{JM}, \quad (2)$$

where $\psi_{jm}(\mathbf{r})$ are single particle wave functions obtained from a (self-consistent) mean-field calculation.

The transfer matrix element and correspondingly the magnitude of the transfer cross section is determined essentially by the overlap of $\psi$ with the initial and final nuclear states leading to the transfer form factors

$$F_{JM}^{(J_A J_B)}(\mathbf{r_1}, \mathbf{r_2}) = \langle J_B M_B | \Psi_{JM}(\mathbf{r_1}, \mathbf{r_2}) | J_A M_A \rangle = \frac{(-)^{J_A+M_A}}{\sqrt{2J+1}} (J_B M_B J_A - M_A | JM)$$
$$\sum_{ik}\left(\phi_{JM}^{(j_i j_k)}(\mathbf{r_1}, \mathbf{r_2}) \langle J_B || P_J^{+}(j_i j_k) || J_A \rangle + (-)^{J_A+J_B+J-M} \tilde{\phi}_{JM}^{(j_i j_k)}(\mathbf{r_1}, \mathbf{r_2}) \langle J_B || P_J(j_i j_k) || J_A \rangle\right) \quad (3)$$

Two kinds of spectroscopic amplitudes are identified:

$$S^{(+)}_{J_B J_A J}(j_i j_k) = \langle J_B || P^+_J(j_i j_k) || J_A \rangle$$
$$S^{(-)}_{J_B J_A J}(j_i j_k) = \langle J_B || P_J(j_i j_k) || J_A \rangle \tag{4}$$

describing the spectroscopic strength for addition and removal of two nucleons.

In QRPA-theory, the final states are obtained by acting on the parent state with the operators

$$\Omega^+_{nJM} = \sum_{ik} \left( x_{nJ}(j_i j_k) Q^+_{JM}(j_i j_k) - y^*_{nJ}(j_i j_k) \tilde{Q}_{JM}(j_i j_k) \right), \tag{5}$$

with the 2QP state operators $Q^+_{JM}(j_i j_k) = \left[\alpha^+_{j_i} \alpha^+_{j_k}\right]_{JM}$ where $\alpha^+_{jm} = u_j a^+_{jm} - v_j \tilde{a}_{jm}$ is a one-quasiparticle operator, obtained from the particle operators by the Bogoliubov-Valatin transformation with $u_j^2 + v_j^2 = 1$. The QRPA eigenenergies $E_n$ and the configuration amplitudes $x_{nJ}$ and $y_{nJ}$, respectively, are solutions of the QRPA eigenvalue problem, see e.g. [16].

In the present context, the parent states are $J^\pi_A = 0^+$ states denoted by $|0\rangle$. Then, $|J_B M_B\rangle = \Omega^+_{J_B M_B}|0\rangle$ and by orthogonality $\Omega_{J_B M_B}|0\rangle = 0$. The latter property allows to express the configuration amplitudes in terms of a commutator relation, e.g.

$$S^{(+)}_{J_B}(j_i j_k) = \langle 0 || \left[\Omega_{nJ_B}, P^+_J(j_i j_k)\right] || 0 \rangle, \tag{6}$$

where we have used a simplified notation by omitting the now superfluous $J_A$. The commutator is easily evaluated and one finds:

$$S^{(+)}_{J_B}(j_i j_k) = u_{j_i} u_{j_k} x^*_{nJ_B}(j_i j_k) + v_{j_i} v_{j_k} y_{nJ_B}(j_i j_k), \tag{7}$$

where we have neglected minor contributions from rescattering terms. Accordingly, the reduced pair-removal amplitude is given by

$$S^{(-)}_{J_B}(j_i j_k) = v_{j_i} v_{j_k} x^*_{nJ_B}(j_i j_k) + u_{j_i} u_{j_k} y_{nJ_B}(j_i j_k). \tag{8}$$

As an important side-remark we emphasize that the transfer form factors, Eq.(3), as a coherent superposition of a number of terms sensitively dependent on the use of consistent phase conventions for all parts of the wave functions, from radial wave functions and spherical harmonics to the conventions used for angular momentum coupling and the definition of reduced matrix elements. Both the magnitude and the phases of the spectroscopic amplitudes and the pair wave functions do matter.

## 4. Transfer cross section results and discussions

As a first step, we have calculated the two-proton stripping and two-neutron pickup cross section angular distributions from the ground state of the projectile and target nuclei. The blue arrows in Figures 2, 3, 4 and 5 represent the coupling schemes adopted for these transfer reactions. The spectroscopic amplitudes used for the direct and the sequential mechanisms in the two-proton and two-neutron transfer reactions are listed in the Tables of the Appendix. The spectroscopic amplitudes used for the one-neutron and one-proton for the target overlaps are not reported here for the sake of space. We performed Coupled Reaction Channel (CRC) calculations using the independent coordinates scheme for the direct transfer (so, the couplings are included in the infinite order). These calculations are referred to as CRC-1. For the sequential transfer, the two-step DWBA was used (so, the coupling among different partitions is considered to the first order).

A further step in our analysis was to consider the couplings of the projectile and target ground state to their low-lying excited states in order to verify the relevance of these couplings on the transfer cross sections. One should have in mind that the projectile/target might be excited before the transfer reaction occurs. In this way, the first excited state of $^{20}Ne_{1.63}(2^+)$ and, according to the vibrational nature of the target, the one-phonon quadrupole $^{116}Cd_{0.514}(2^+)$ and the two-phonon quadrupole states of $^{116}Cd$ [$^{116}Cd_{1.213}(2^+)$, $^{116}Cd_{1.219}(4^+)$ and $^{116}Cd_{1.283}(0^+)$] were included in the coupling scheme. These calculations are referred to as CRC-2 and the included states are shown in Figure 2 and 4. The coupling with the collective states of the projectile and target is obtained by deforming the Coulomb and nuclear potentials. The quadrupole deformations $\beta_2 = 0.720$ for the $^{20}Ne$ and $\beta_2 = 0.135$ for $^{116}Cd$ were taken from the Raman's systematics [70]. These parameters are important to calculate the intrinsic reduced electric quadrupole transition strength and the deformation length corresponding to the Coulomb and nuclear deformations.

The sequential transfer reaction mechanism was treated also within the Coupled Channel Born Approximation (CCBA), in which the couplings in the entrance partition are considered to infinite orders and those among the partitions to the first order [71]. In Figures 3 and 5 we omitted the couplings with the two-phonon quadrupole states of the $^{116}Cd$ in the entrance partition, besides other excited states in the final partition, for better readability of the scheme. In two-nucleon sequential transfer involving a heavy nucleus, the density of states of the odd intermediate nuclei is too high and it is impracticable to include all of them in the coupling scheme. Therefore, we only included the levels in the range of 0-2 MeV for single-neutron states and 0-2.4 MeV for single-proton ones.

### 4.1 Analysis of the two-neutron pickup $^{116}$Cd($^{20}$Ne,$^{22}$Ne)$^{114}$Cd reaction cross section

In this sub-section, we show the results obtained for the $^{116}$Cd($^{20}$Ne,$^{22}$Ne)$^{114}$Cd reaction. In Table 6, the theoretical results for transitions to the $^{22}$Ne$_{gs}$(0$^+$) + $^{114}$Cd$_{gs}$(0$^+$) and $^{22}$Ne$_{gs}$(0$^+$) + $^{114}$Cd$_{0.558}$(2$^+$) exit channels are compared with the corresponding experimental values. For the direct transfer the spectroscopic amplitudes were derived from shell-model for projectile overlaps (Table A1) and from both shell model and microscopic IBM-2 for target overlaps (Table A3 and A4, respectively). In the sequential transfer process, the spectroscopic amplitudes for the projectile overlaps (Table A2) were derived by using the *p-sd-mod* interaction and for the target overlaps by using the *jj45pna* one. All the two-neutron transfer cross sections were integrated upon the angular range 4.5° ≤ $\theta_{lab}$ ≤ 14.5° in the laboratory frame for consistency with the experiment. The IC CRC-1 and Seq DWBA correspond to the results obtained using the couplings represented by the blue arrows in Figures 2 and 3. The CRC-2 and Seq CCBA correspond to the full coupling shown in the same figures, i.e., they also take into account the couplings with the inelastic states in the entrance partition.

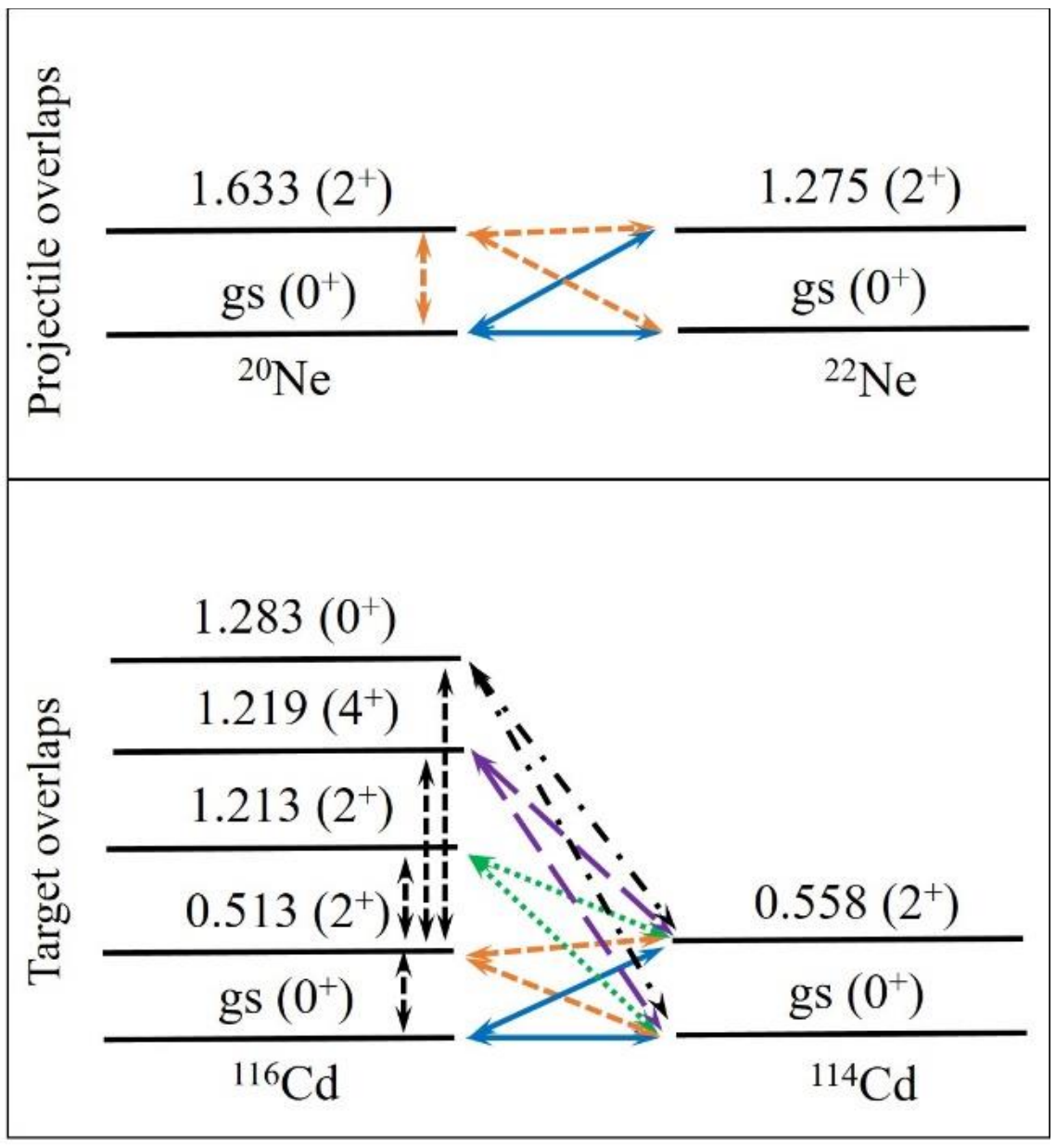

**Figure 2**. Couplings schemes for projectile and target overlaps considered in the two-neutron direct transfer calculations.

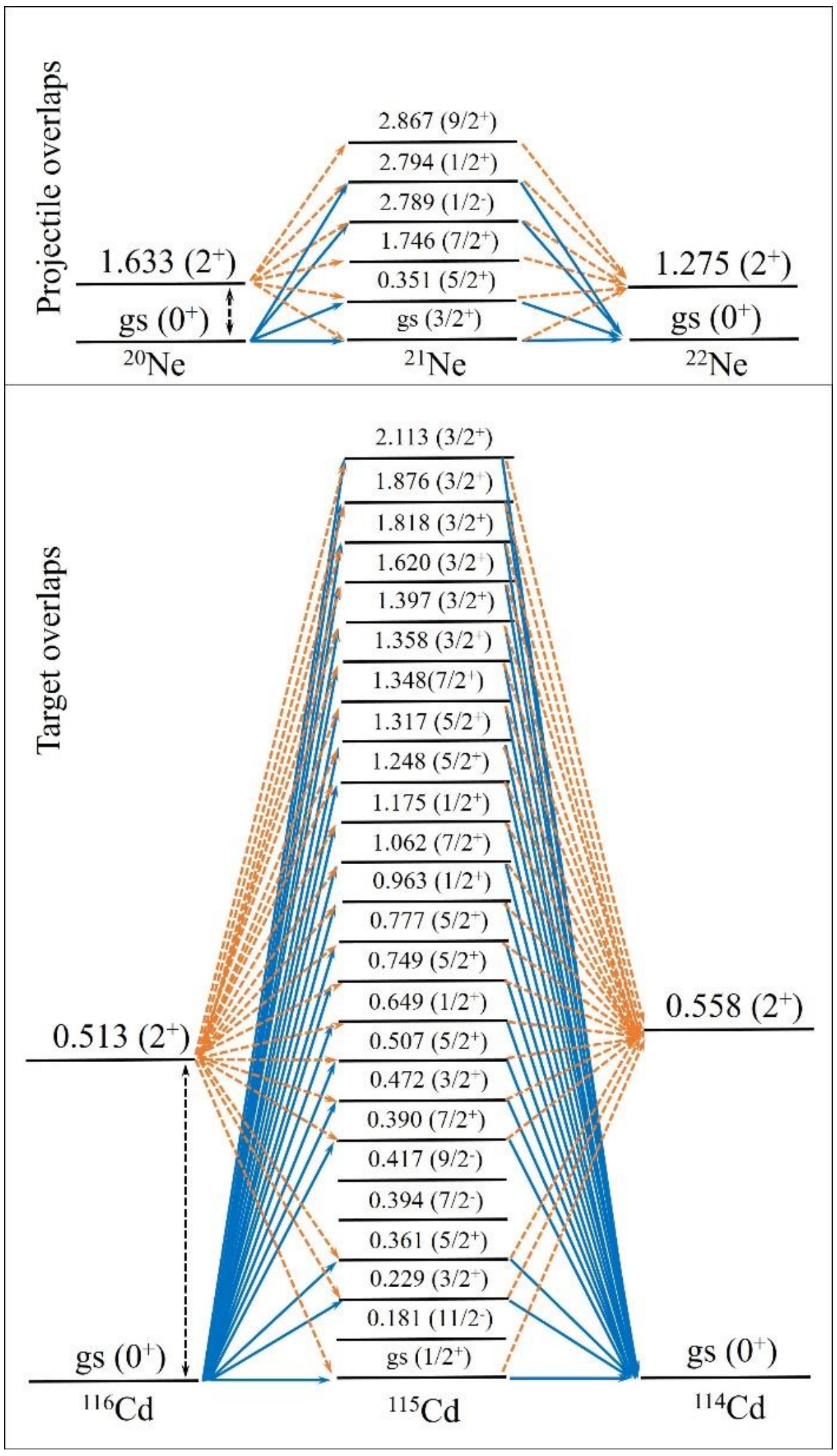

**Figure 3**. Couplings schemes for projectile and target overlaps considered in the two-neutron sequential transfer calculations. The couplings with the two-phonon states of the $^{116}$Cd are omitted.

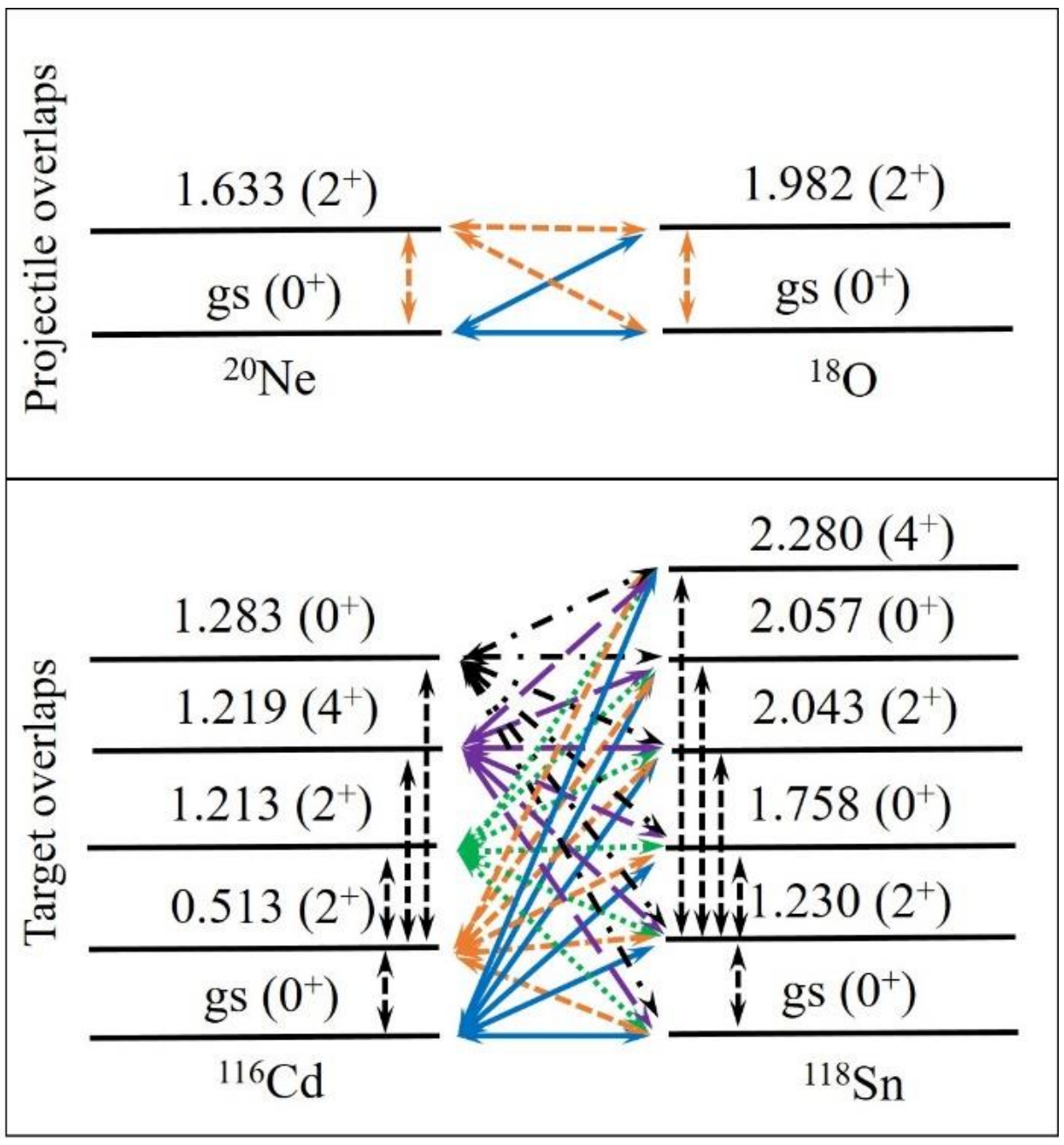

**Figure 4**. Couplings schemes for the projectile and target overlaps considered in the two-proton direct transfer calculations.

First, one can compare the theoretical cross sections for simultaneous transfer (IC model) obtained from CRC-1 and CRC-2 with the experimental ones. The theoretical cross section for the $^{22}Ne_{gs}(0^+)$ + $^{114}Cd_{gs}(0^+)$ channel, either using the amplitudes from SM or IBM-2 models, do not strongly deviate from the experimental data within the error bar, although the value obtained by using the IBM-2 amplitudes is more than twice the one from SM calculation. This discrepancy comes from the inclusion of the $1h_{11/2}$ orbital in the IBM-2 calculations, which is missing in the model space of the SM. For the $^{22}Ne_{gs}(0^+)$ + $^{114}Cd_{0.514}(2^+)$ channel, the SM calculations agree with the data either using the CRC-1 or the CRC-2 approach. In the IBM-2 calculations, even if the inclusion of the inelastic couplings increases the resulting cross section (IC CRC-2 value for IBM-2 in Table 6) still it underestimates significantly the experimental value.

Then, the two neutrons are assumed to be transferred in a sequential way. As it can be seen in Table 6, the result for the $^{22}Ne_{gs}(0^+)$ + $^{114}Cd_{gs}(0^+)$ channel (Seq DWBA) gives a reasonable description of the experimental data. On the other hand, the theoretical prediction for the $^{22}Ne_{gs}(0^+)$ + $^{114}Cd_{0.558}(2^+)$ channel overestimates a little bit the experimental value. The agreement between theory and experiment is improved when the couplings with the inelastic states in the entrance partition are explicitly considered in the sequential transfer calculation (Seq CCBA).

In principle, a coherent sum between the CRC and the sequential calculations would be needed, but this procedure requires a fit on the experimental cross section angular distributions, which were not extracted from the data due to the poor statistics. The algebraic sum of CRC and sequential can be considered in this case an estimate of the total theoretical cross section, keeping in mind that it can be an overestimation (or even underestimation) depending on the relative phases. The same argument holds for the two-proton transfer calculations discussed in the next subsection.

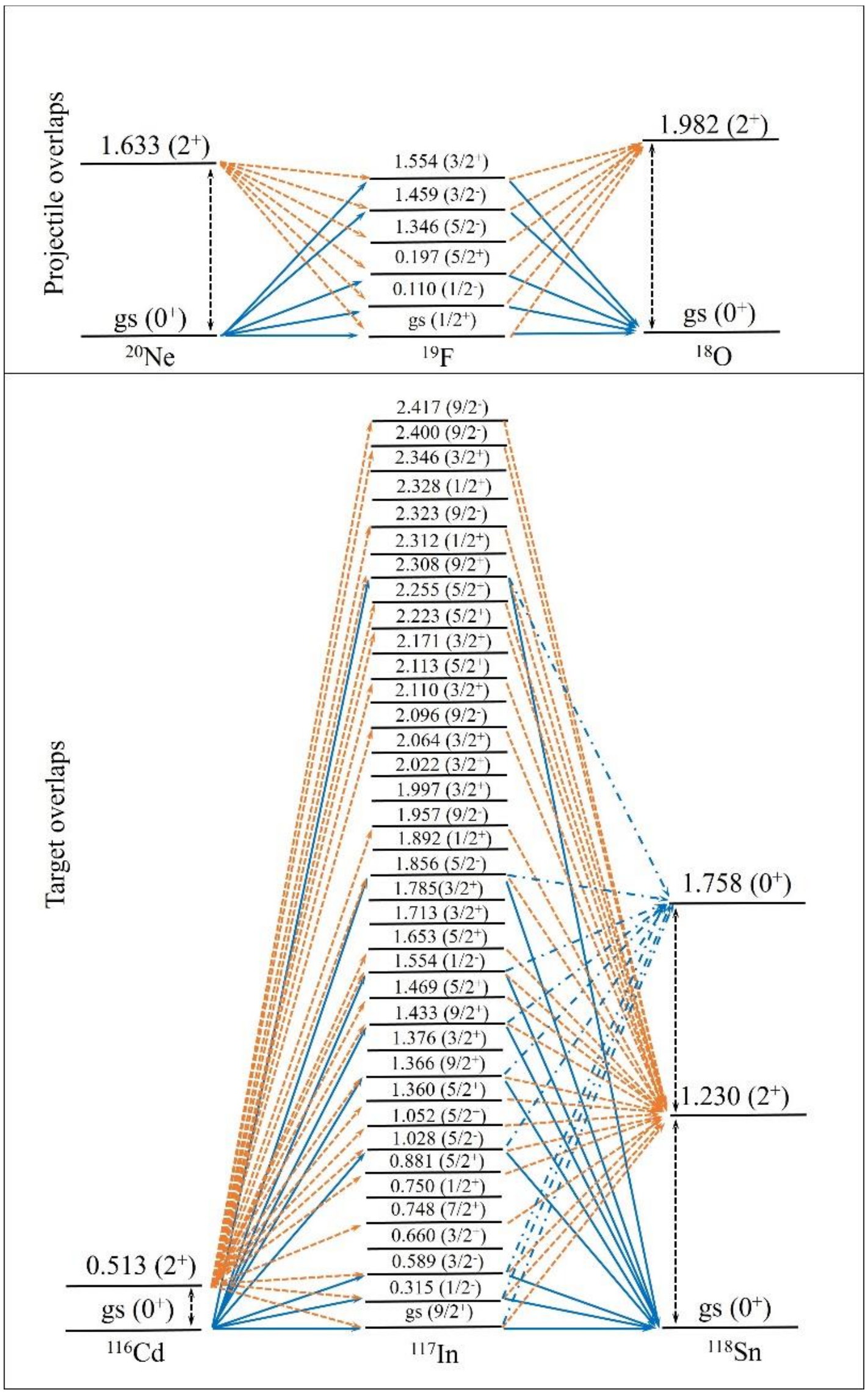

**Figure 5**. Couplings schemes for the projectile and target overlap considered in the two-proton sequential transfer calculations. Here, the couplings with the two-phonon states of the $^{116}$Cd and for excited states above 1.758 MeV of the $^{118}$Sn in the two-proton transfer are omitted.

**Table 6**. Comparison between experimental and theoretical integrated cross sections for the two-neutron pickup (for 4.5° < $\theta_{lab}$ < 14.5°). The amplitude for the projectile overlaps were derived by SM calculations using the *p-sd-mod* interaction. For the target overlaps, the results using the *jj45pna* in the SM approach and the microscopic IBM-2 are reported.

| **Final Channel** | **Cross Sections (nb)** | | | | | | |
|---|---|---|---|---|---|---|---|
| | **Exp.** | **Theory** | | | | | |
| | | **SA-shell model – *psdmod* + *jj45pna*** | | | | **SA-IBM-2** | |
| | | IC CRC-1 | Seq DWBA | IC CRC-2 | Seq CCBA | IC CRC-1 | IC CRC-2 |
| $^{22}Ne_{gs}(0^+)$ + $^{114}Cd_{gs}(0^+)$ | 450 ± 200 | 251 | 613 | 209 | 427 | 689 | 572 |
| $^{22}Ne_{gs}(0^+)$ + $^{114}Cd_{0.558}(2^+)$ | 420 ± 190 | 313 | 721 | 314 | 636 | 8 | 28 |

It would be interesting to study the behaviour of the angular distributions, which could provide more details of the reaction processes studied in the present work. Looking at the calculated two-neutron transfer angular distribution in Figure 6, a pronounced bell-shape peaked at the grazing angle (~ 13°) is observed resembling what found in refs. [72] [27]. This effect corresponds to the trade-off of the strong absorption experienced for impact parameters smaller than $R = R_T + R_P$, with $R_T$ and $R_P$ the target and projectile radii, respectively, and the range of the attractive nuclear interaction of the host nucleus felt by the transferred neutrons. As a result, the transfer process is more likely to occur for impact parameters within a narrow range around $R$ for which the grazing condition is established [73].

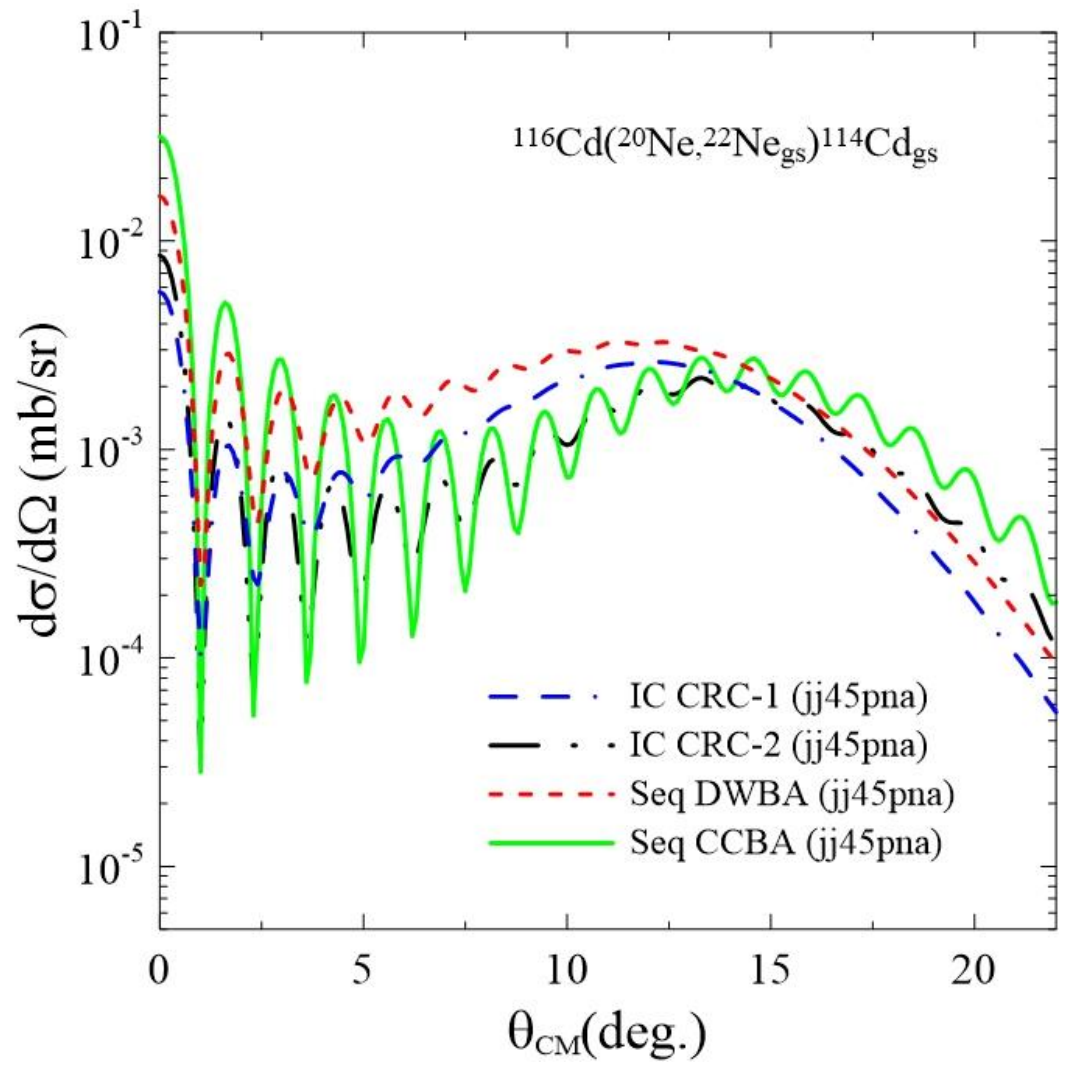

**Figure 6.** Theoretical angular distributions obtained for the $^{116}Cd(^{20}Ne,^{22}Ne_{g.s.})^{114}Cd_{g.s.}$ two-neutron transfer reaction. The IC CRC-1 (blue double-dashed-dotted line) and Seq DWBA (red dashed) consider only the ground state of the projectile and target nuclei in the entrance partition. The IC CRC-2 (black dashed-double-dotted line) and Seq CCBA (green line) consider the full coupling sketched in Figures 2 and 3.

From Figure 6, one clearly sees that the one-step process competes with the sequential one, showing that the pairing correlations between the two neutrons in $^{116}Cd_{g.s.}$ and $^{20}Ne_{g.s.}$ play a role in the two-neutron transfer process. This is in agreement with the results recently obtained for the ($^{18}O$,$^{16}O$) reaction on light and medium- mass systems, especially for states with low-collectivity (small reduced quadrupole transition probabilities B(E2)) [27]. In Figure 6, the effect of the inelastic states in the entrance partition is also emphasized. The inclusion of these collective couplings in the system of coupled equations produces a phase shift on the angular distribution and increases the magnitude of the oscillations. The addition of these couplings provides also a decrease in the transfer cross sections at the very forward angles and an increase above the grazing angle.

### 4.2 Analysis of two-proton stripping $^{116}$Cd($^{20}$Ne,$^{18}$O)$^{118}$Sn reaction cross section

Here we show the theoretical analysis for the two-proton transfer cross sections adopting the same procedure used in the two-neutron transfer calculations. Since the application of this framework is discussed for the first time here in the two-proton transfer case, we decided to treat the nuclear structure information with a supplementary approach. For the projectile, the spectroscopic amplitudes were derived using the *p-sd-mod* interaction (Table A1 and A2) in the shell model calculations. For the target overlaps, the two-proton amplitudes were obtained from the shell-model calculations with the *jj45pna* and *88Sr45* (Table A5) phenomenological interactions, from the microscopic Interacting Boson Model-2 (IBM-2) (Table A6) and from the QRPA (Table A7). The coupling schemes considered for the projectile and target overlaps are sketched in Figures 4 and 5.

The coupling between ground and excited states were introduced in the initial and final partition by deforming the optical potential. The strength coefficient $N_I$ = 0.5 was considered in the imaginary part of the optical potential in the entrance and final partitions since we are explicitly considering the couplings with relevant excited states. In particular, as both $^{18}$O and $^{118}$Sn nucleus have a closed shell for protons, the proton excitation in those nuclei could be less likely to occur which would justify the optical potential a little bit less absorptive on the nuclear surface (the usual strength coefficient used in the literature is $N_I$ = 0.6 [25] [26] [27] [28] [29] [30] [31]).

In Table 7 we compare the calculated two-proton cross sections for the $^{18}O_{gs}(0^+)$ + $^{118}Sn_{gs}(0^+)$ and $^{18}O_{gs}(0^+)$ + $^{118}Sn_{1.229}(2^+)$ transitions with the experimental values. The cross sections were integrated into the angular range 4° < $\theta_{lab}$ < 14° to be compared with the experimental results.

We start our two-proton transfer analysis comparing the results obtained for the two-proton transfer in which both valence protons are simultaneously transferred (IC scheme) and using the *jj45pna* interaction in the SM. In this case, when no coupling with the inelastic states is considered in the initial partition, the predicted two-proton transfer cross sections are slightly smaller than the data for the $^{18}O_{gs}(0^+)$ + $^{118}Sn_{gs}(0^+)$ channel (see IC CRC-1 results in Table 7). The agreement is reached after the inclusion of the inelastic couplings (IC CRC-2). The results are more critical for the $^{18}O_{gs}(0^+)$ + $^{118}Sn_{1.229}(2^+)$ channel, for which the calculated values are smaller than the experimental data without or with the inclusion of the couplings.

For the sequential two-proton transfer, the comparison between the Seq DWBA and Seq CCBA results shows the importance of considering the couplings with inelastic states of the $^{20}$Ne and $^{116}$Cd nuclei. Again the theoretical CCBA two-proton transfer calculations describe very well the experimental cross section for the $^{18}O_{gs}(0^+)$ + $^{118}Sn_{gs}(0^+)$ channel, while for the first excited state they underestimate the experimental value.

**Table 7**. Comparison between experimental and theoretical integrated cross sections corresponding to the two-proton stripping (for 4° < $\theta_{lab}$ < 14°) transfer processes. The amplitudes for the projectile overlaps were derived by shell model calculation using the *p-sd*-mod interaction. For the target overlaps, the results using the jj45pna and 88Sr45 interactions within the SM, microscopic IBM-2 and the QRPA are reported.

| Final Channel | | Cross Sections (nb) | | | | | | | | |
|---|---|---|---|---|---|---|---|---|---|---|
| | Exp. | Theory | | | | | | | | |
| | | SA-shell model *p-sd-mod + jj45pna* int. | | | | SA-shell model *p-sd-mod + 88Sr45* | | SA IBM-2 | | SA QRPA |
| | | IC CRC-1 | Seq DWBA | IC CRC-2 | Seq CCBA | IC CRC-2 | Seq CCBA | IC CRC-1 | IC CRC-2 | IC CRC-1 |
| $^{18}O_{gs}(0^+) + {}^{118}Sn_{gs}(0^+)$ | 40 ± 15 | 22 | 19.1 | 30.9 | 52.1 | 39.5 | 88.5 | 32.7 | 23.1 | 19 |
| $^{18}O_{gs}(0^+) + {}^{118}Sn_{1.229}(2^+)$ | 140 ± 60 | 5.3 | 1.6 | 26.9 | 39.8 | 52.7 | 106.3 | 3.1 | 2.6 | 55 |

One can argue that the model space considered for the valence protons might be not enough, since the higher orbits could be important to describe the structure of the heavier nuclei. To investigate this aspect, in a second step we included the $1g_{7/2}$ and $2d_{5/2}$ orbitals in the model space of the protons using the *88Sr45* interaction to calculate the spectroscopic amplitudes for the target overlaps. Indeed, as one can see in Table 7, the obtained results for both transitions are now in better agreement with the experimental data, showing the importance of the added orbitals especially for the $^{18}O_{gs}(0^+) + {}^{118}Sn_{1.229}(2^+)$ transition.

As we observed for the $^{22}Ne_{gs}(0^+) + {}^{114}Cd_{gs}(0^+)$ transition, the results with microscopic IBM-2 in the two-proton case are smaller than those with the SM. This is a general trend already observed by us in refs. [27] [58] that microscopic IBM-2 results are smaller than shell model ones. Looking at Table 7, we see that the IBM-2 integrated cross section for the $^{18}O_{gs}(0^+) + {}^{118}Sn_{gs}(0^+)$ channel is inside the experimental value, whereas the $^{18}O_{gs}(0^+) + {}^{118}Sn_{1.229}(2^+)$ transition value underestimates the data both in the CRC-1 and CRC-2 calculations. One reason for that is the lack of the spectroscopic amplitudes for $^{116}Cd$ to $^{118}Sn(4^+)$ since the transfer operator that brings angular momenta equal to 4 has not been developed in this formalism.

For the two-proton transfer case we explored also the QRPA approach to derive the spectroscopic amplitudes and the results are in acceptable agreement with the experimental data for both transitions.

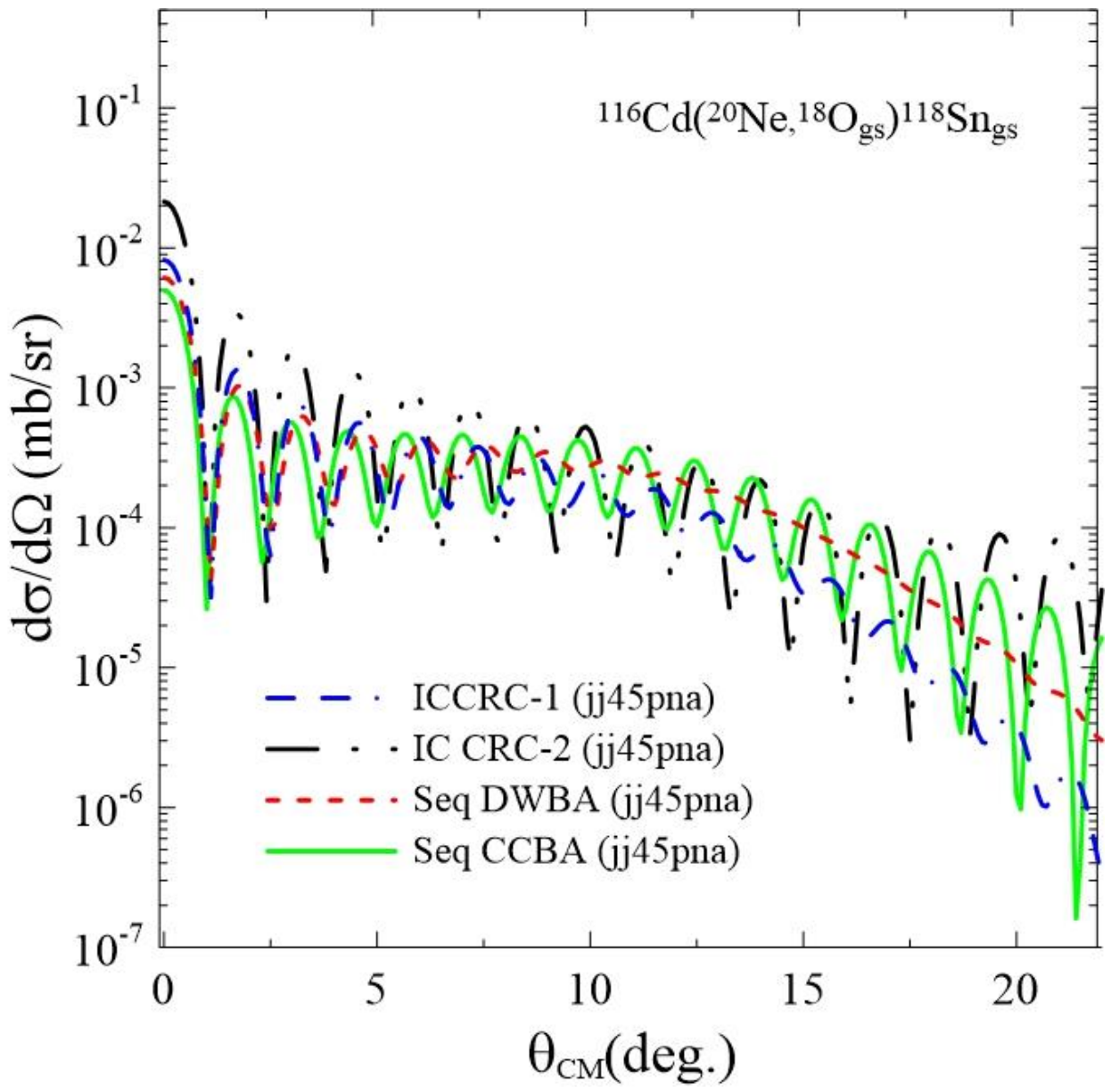

**Figure 7.** Theoretical angular distributions obtained for the $^{116}Cd(^{20}Ne,^{18}O_{gs})^{118}Sn_{gs}$ two-proton stripping transfer reactions. The IC CRC-1 (blue double-dashed-dotted line) and Seq DWBA (red dashed) consider only the ground state of the projectile and target nuclei in the entrance partition. The IC CRC-2 (black dashed-double-dotted line) and Seq CCBA (green line) consider the full coupling sketched in Figures 4 and 5.

In Figure 7, the theoretical angular distributions associated with the two-proton transfer reactions are shown, where both $^{18}O$ and $^{118}Sn$ nuclei are found in their ground states. The two-proton transfer angular distribution shows a somewhat attenuated bell-shape behaviour when compared to the two-neutron transfer one (see Figure 6). On the other hand, the oscillating pattern is extended up to larger scattering angles. This suggests a weaker absorption effect experienced by the two transferred protons in the nuclear field of the target, probably due to the Coulomb repulsion between the two protons and the host core. The one-step process competes with the sequential one showing that the correlations between valence protons in $^{20}Ne_{g.s.}$ wave function are relevant. This characteristic has also been observed in the two-neutron transfer. It is important to notice that the pairing correlation remains relevant in the case of two-proton transfer even if the repulsion between the core and the two protons and between them might attenuate its effect. This conclusion suggests a symmetric behaviour of two-proton and two-neutron transfer reactions, stemming from charge symmetry of nuclear forces and preserving the reaction dynamics.

Similarly to the case of the two-neutron transfer, the inclusion of the inelastic states in the entrance partition produces a phase-shift and increases the magnitude of the oscillations (see Figure 7).

## 5. Summary and conclusions

In the present work, the cross sections for specific final channels corresponding to the two-neutron pickup $^{116}$Cd($^{20}$Ne,$^{22}$Ne)$^{114}$Cd and two-proton stripping $^{116}$Cd($^{20}$Ne,$^{18}$O)$^{118}$Sn reactions were analysed. The experiment was performed at the INFN-LNS laboratory in Catania in the framework of the NUMEN project where the K800 Superconducting Cyclotron beam has accelerated the $^{20}$Ne$^{4+}$ beam at 306 MeV incident energy.

The theoretical two-nucleon transfer cross sections were calculated considering that both valence nucleons may be transferred directly from the initial to the final partition, or they may be transferred one by one, passing through an intermediate partition by a sequential process.

In the $^{116}$Cd($^{20}$Ne,$^{22}$Ne)$^{114}$Cd two-neutron transfer reaction, according to theoretical predictions, the final measured channel $^{22}$Ne$_{gs}$(0$^+$)+$^{114}$Cd$_{gs}$(0$^+$) and $^{22}$Ne$_{gs}$(0$^+$)+$^{114}$Cd$_{0.558}$(2$^+$) have been populated through a competition between two-neutron simultaneous or sequential transfer processes. For the two-proton transfer process, while the direct and sequential processes compete with each other to populate the $^{18}$O$_{gs}$(0$^+$) +$^{118}$Sn$_{gs}$(0$^+$) channel, the sequential process is dominant over the direct one to populating the $^{118}$Sn$_{1.23}$(2$^+$) excited state. Also, an important contribution from the inelastic couplings in the entrance partition is found.

From the structure calculation side, the spectroscopic amplitudes for the target overlaps were derived from the microscopic Shell Model, Interacting Boson Model-2 and Quasiparticle Random Phase Approximation approaches. In the two-neutron transfer case, the same model space was used in the SM and IBM-2 approaches obtaining a good agreement with the experimental cross sections for the two analysed transitions. The description of the cross section in the two-proton transfer case was more difficult. Two different model spaces for the valence protons were considered in the effective Hamiltonian, since it was necessary to include higher orbits ($1g_{7/2}$ and $2d_{5/2}$) in order to obtain the agreement between the experiment and theory for the transition to the first 2$^+$ excited state of $^{118}$Sn. Other approaches were tested in this case, IBM-2 and QRPA, obtaining in both cases a reasonable description of the direct mechanism.

To summarize, a satisfactory description of the two-particle transfer reaction was obtained in the current work for which the microscopic treatment of reaction and nuclear structure aspects was of central importance. These results are a confirmation of the validity of this approach for the two-neutron transfer reactions, which were already studied in the same framework and with lighter nuclei. Moreover, this approach represents a very promising tool for the two-proton transfer reactions between heavy-nuclei, which are analysed here for the first time.

An important application of this work will be in the analysis of double charge exchange reactions, for which these two-particle transfer reactions are the first (in the case of the direct transfer) or the first two steps (in the case of sequential transfer) of the multi-nucleon transfer reactions that might compete with the direct meson exchange mechanism. The framework applied in this work could be safely used to predict the multi-nucleon transfer cross section leading to the same DCE channels explored, for example, in the NUMEN project [13], since for some possible steps there are no experimental information.

**Acknowledgements**

This project has received funding from the European Research Council (ERC) under the European Union's Horizon 2020 research and innovation programme (grant agreement No 714625). The Brazilian authors acknowledge partial financial support from CNPq, FAPERJ, and CAPES and from INCT-FNA (Instituto Nacional de Ciência e Tecnologia – Física Nuclear e Aplicações), research project 464898/2014-5. The Mexican authors acknowledge partial financial support from CONACyT 299073 and PAPIIT-DGAPA IN107820 and AG101120. The authors wish to thank A. Gargano for fruitful discussions and suggestions. The authors wish to acknowledge the INFN-LNS acceleration division staff for the excellent job done in setting up the beam.

# Appendix

**Table A1.** Two-proton and two-neutron spectroscopic amplitudes for the projectile overlaps using the *p-sd-mod* interaction.

| Initial state | $j_1j_2$ | $J_{12}$ | Final state | S. A. |
|---|---|---|---|---|
| $^{20}Ne_{gs}(0^+)$ | $1p_{3/2}\,1p_{3/2}$ | 0 | $^{18}O_{gs}(0^+)$ | -0.07228 |
| | $1p_{1/2}\,1p_{1/2}$ | | | -0.20700 |
| | $1d_{5/2}\,1d_{5/2}$ | | | 0.53508 |
| | $1d_{3/2}\,1d_{3/2}$ | | | 0.19785 |
| | $2s_{1/2}\,2s_{1/2}$ | | | 0.23605 |
| | $1p_{3/2}\,1p_{3/2}$ | 2 | $^{18}O_{1.98}(2^+)$ | -0.00441 |
| | $1p_{3/2}\,1p_{1/2}$ | | | 0.01459 |
| | $1d_{5/2}\,1d_{5/2}$ | | | -0.33083 |
| | $1d_{5/2}\,1d_{3/2}$ | | | 0.16698 |
| | $1d_{5/2}\,2s_{1/2}$ | | | -0.35539 |
| | $1d_{3/2}\,1d_{3/2}$ | | | -0.13504 |
| | $1d_{3/2}\,2s_{1/2}$ | | | -0.19493 |
| $^{20}Ne_{1.63}(2^+)$ | $1p_{3/2}\,1p_{3/2}$ | 2 | $^{18}O_{gs}(0^+)$ | 0.00084 |
| | $1p_{3/2}\,1p_{1/2}$ | | | -0.00205 |
| | $1d_{5/2}\,1d_{5/2}$ | | | 0.27233 |
| | $1d_{5/2}\,1d_{3/2}$ | | | -0.10637 |
| | $1d_{5/2}\,2s_{1/2}$ | | | 0.31791 |
| | $1d_{3/2}\,1d_{3/2}$ | | | 0.08214 |
| | $1d_{3/2}\,2s_{1/2}$ | | | 0.16589 |
| | $1p_{3/2}\,1p_{3/2}$ | 0 | $^{18}O_{1.98}(2^+)$ | 0.04284 |
| | $1p_{1/2}\,1p_{1/2}$ | | | 0.14392 |
| | $1d_{5/2}\,1d_{5/2}$ | | | -0.40846 |
| | $1d_{3/2}\,1d_{3/2}$ | | | -0.13797 |
| | $2s_{1/2}\,2s_{1/2}$ | | | -0.21641 |
| $^{20}Ne_{gs}(0^+)$ | $1p_{3/2}\,1p_{3/2}$ | 0 | $^{22}Ne_{gs}(0^+)$ | -0.05578 |
| | $1p_{1/2}\,1p_{1/2}$ | | | -0.18665 |
| | $1d_{5/2}\,1d_{5/2}$ | | | 0.74430 |
| | $1d_{3/2}\,1d_{3/2}$ | | | 0.19033 |
| | $2s_{1/2}\,2s_{1/2}$ | | | 0.18162 |
| | $1p_{3/2}\,1p_{3/2}$ | 2 | $^{22}Ne_{1.275}(2^+)$ | -0.00117 |
| | $1p_{3/2}\,1p_{1/2}$ | | | -0.02476 |
| | $1d_{5/2}\,1d_{5/2}$ | | | 0.19478 |
| | $1d_{5/2}\,1d_{3/2}$ | | | -0.16369 |
| | $1d_{5/2}\,2s_{1/2}$ | | | 0.14327 |
| | $1d_{3/2}\,1d_{3/2}$ | | | 0.03964 |
| | $1d_{3/2}\,2s_{1/2}$ | | | 0.07193 |
| $^{20}Ne_{1.63}(2^+)$ | $1p_{3/2}\,1p_{3/2}$ | 2 | $^{22}Ne_{gs}(0^+)$ | 0.00158 |
| | $1p_{3/2}\,1p_{1/2}$ | | | -0.10601 |
| | $1d_{5/2}\,1d_{5/2}$ | | | -0.28319 |
| | $1d_{5/2}\,1d_{3/2}$ | | | -0.21805 |
| | $1d_{5/2}\,2s_{1/2}$ | | | -0.13975 |
| | $1d_{3/2}\,1d_{3/2}$ | | | -0.02115 |
| | $1d_{3/2}\,2s_{1/2}$ | | | -0.03537 |
| | $1p_{3/2}\,1p_{3/2}$ | 0 | $^{22}Ne_{1.275}(2^+)$ | -0.04660 |
| | $1p_{1/2}\,1p_{1/2}$ | | | -0.17062 |
| | $1d_{5/2}\,1d_{5/2}$ | | | 0.58410 |
| | $1d_{3/2}\,1d_{3/2}$ | | | 0.14152 |
| | $2s_{1/2}\,2s_{1/2}$ | | | 0.08920 |

**Table A2.** One-proton and one-neutron spectroscopic amplitudes for the projectile overlaps using the *p-sd-mod* interaction.

| Initial state | $nl_j$ | Final state | S. A. |
|---|---|---|---|
| $^{20}\mathrm{Ne}_{gs}(0^+)$ | $1d_{3/2}$ | $^{21}\mathrm{Ne}_{gs}(3/2^+)$ | -0.15863 |
| | $1d_{5/2}$ | $^{21}\mathrm{Ne}_{0.351}(5/2^+)$ | 0.75401 |
| | $1p_{1/2}$ | $^{21}\mathrm{Ne}_{2.789}(1/2^-)$ | 0.34146 |
| | $2s_{1/2}$ | $^{21}\mathrm{Ne}_{2.794}(1/2^+)$ | 0.82780 |
| $^{20}\mathrm{Ne}_{1.63}(2^+)$ | $2s_{1/2}$ | $^{21}\mathrm{Ne}_{gs}(3/2^+)$ | 0.17092 |
| | $1d_{3/2}$ | | 0.20139 |
| | $1d_{5/2}$ | | -0.93303 |
| | $2s_{1/2}$ | $^{21}\mathrm{Ne}_{0.351}(5/2^+)$ | -0.17532 |
| | $1d_{3/2}$ | | -0.13304 |
| | $1d_{5/2}$ | | -0.25974 |
| | $1d_{3/2}$ | $^{21}\mathrm{Ne}_{1.746}(7/2^+)$ | -0.04225 |
| | $1d_{5/2}$ | | 0.83879 |
| | $1p_{3/2}$ | $^{21}\mathrm{Ne}_{2.789}(1/2^-)$ | -0.16667 |
| | $1d_{3/2}$ | $^{21}\mathrm{Ne}_{2.794}(1/2^+)$ | 0.01675 |
| | $1d_{5/2}$ | | -0.49360 |
| | $1d_{5/2}$ | $^{21}\mathrm{Ne}_{2.867}(9/2^+)$ | -0.64530 |
| $^{22}\mathrm{Ne}_{gs}(0^+)$ | $1d_{3/2}$ | $^{21}\mathrm{Ne}_{gs}(3/2^+)$ | -0.32481 |
| | $1d_{5/2}$ | $^{21}\mathrm{Ne}_{0.351}(5/2^+)$ | 1.50037 |
| | $1p_{1/2}$ | $^{21}\mathrm{Ne}_{2.789}(1/2^-)$ | -0.83957 |
| | $2s_{1/2}$ | $^{21}\mathrm{Ne}_{2.794}(1/2^+)$ | 0.33823 |
| $^{22}\mathrm{Ne}_{1.275}(2^+)$ | $2s_{1/2}$ | $^{21}\mathrm{Ne}_{gs}(3/2^+)$ | 0.08719 |
| | $1d_{3/2}$ | | 0.08764 |
| | $1d_{5/2}$ | | 0.93329 |
| | $2s_{1/2}$ | $^{21}\mathrm{Ne}_{0.351}(5/2^+)$ | 0.23017 |
| | $1d_{3/2}$ | | 0.27006 |
| | $1d_{5/2}$ | | 0.41915 |
| | $1d_{3/2}$ | $^{21}\mathrm{Ne}_{1.746}(7/2^+)$ | 0.16855 |
| | $1d_{5/2}$ | | -0.59655 |
| | $1p_{3/2}$ | $^{21}\mathrm{Ne}_{2.789}(1/2^-)$ | -0.09145 |
| | $1d_{3/2}$ | $^{21}\mathrm{Ne}_{2.794}(1/2^+)$ | 0.10989 |
| | $1d_{5/2}$ | | 0.19015 |
| $^{20}\mathrm{Ne}_{gs}(0^+)$ | $2s_{1/2}$ | $^{19}\mathrm{F}_{gs}(1/2^+)$ | -0.61964 |
| | $1p_{1/2}$ | $^{19}\mathrm{F}_{0.110}(1/2^-)$ | 1.17339 |
| | $1d_{5/2}$ | $^{19}\mathrm{F}_{0.197}(5/2^+)$ | -1.17502 |
| | $1p_{3/2}$ | $^{19}\mathrm{F}_{1.459}(3/2^-)$ | -0.33486 |
| | $1d_{3/2}$ | $^{19}\mathrm{F}_{1.554}(3/2^+)$ | -0.58241 |
| $^{20}\mathrm{Ne}_{1.63}(2^+)$ | $1d_{3/2}$ | $^{19}\mathrm{F}_{gs}(1/2^+)$ | -0.29449 |
| | $1d_{5/2}$ | | -0.58382 |
| | $1p_{3/2}$ | $^{19}\mathrm{F}_{0.110}(1/2^-)$ | 0.08880 |
| | $2s_{1/2}$ | $^{19}\mathrm{F}_{0.197}(5/2^+)$ | -0.52774 |
| | $1d_{3/2}$ | | -0.17698 |
| | $1d_{5/2}$ | | -0.61711 |
| | $1p_{1/2}$ | $^{19}F_{1.346}(5/2^-)$ | -0.88371 |
| | $1p_{3/2}$ | | -0.10150 |
| | $1p_{1/2}$ | $^{19}F_{1.459}(3/2^-)$ | 0.73144 |
| | $1p_{3/2}$ | | -0.08822 |
| | $2s_{1/2}$ | $^{19}F_{1.554}(3/2^+)$ | 0.37313 |
| | $1d_{3/2}$ | | -0.24621 |
| | $1d_{5/2}$ | | 0.26002 |
| $^{18}O_{gs}(0^+)$ | $2s_{1/2}$ | $^{19}F_{gs}(1/2^+)$ | -0.55389 |
| | $1p_{1/2}$ | $^{19}F_{0.110}(1/2^-)$ | -0.24680 |
| | $1d_{5/2}$ | $^{19}F_{0.197}(5/2^+)$ | -0.66286 |
| | $1p_{3/2}$ | $^{19}F_{1.459}(3/2^-)$ | 0.01100 |
| | $1d_{3/2}$ | $^{19}F_{1.554}(3/2^+)$ | -0.42310 |
| $^{18}O_{1.98}(2^+)$ | $1d_{3/2}$ | $^{19}F_{gs}(1/2^+)$ | -0.28172 |
| | $1d_{5/2}$ | | 0.58600 |
| | $1p_{3/2}$ | $^{19}F_{0.110}(1/2^-)$ | -0.03101 |
| | $2s_{1/2}$ | $^{19}F_{0.197}(5/2^+)$ | 0.31290 |
| | $1d_{3/2}$ | | -0.15663 |
| | $1d_{5/2}$ | | 0.42572 |
| | $1p_{1/2}$ | $^{19}F_{1.346}(5/2^-)$ | -0.13790 |
| | $1p_{3/2}$ | | -0.01873 |
| | $1p_{1/2}$ | $^{19}F_{1.459}(3/2^-)$ | -0.16510 |
| | $1p_{3/2}$ | | 0.00199 |
| | $2s_{1/2}$ | $^{19}F_{1.554}(3/2^+)$ | 0.35505 |
| | $1d_{3/2}$ | | 0.31833 |
| | $1d_{5/2}$ | | 0.31429 |

**Table A3.** Two-neutron spectroscopic amplitudes for the target overlaps using the *jj45pna* interaction.

| Initial state | $j_1j_2$ | $J_{12}$ | Final state | S. A. |
|---|---|---|---|---|
| $^{116}Cd_{gs}(0^+)$ | $1g_{7/2}\,1g_{7/2}$ | 0 | $^{114}Cd_{gs}(0^+)$ | -0.5960 |
| | $2d_{5/2}\,2d_{5/2}$ | | | -0.2829 |
| | $2d_{3/2}\,2d_{3/2}$ | | | -0.6415 |
| | $3s_{1/2}\,3s_{1/2}$ | | | -0.3341 |
| | $1g_{7/2}\,1g_{7/2}$ | 2 | $^{114}Cd_{0.558}(2^+)$ | 0.6187 |
| | $1g_{7/2}\,2d_{5/2}$ | | | 0.1006 |
| | $1g_{7/2}\,2d_{3/2}$ | | | 0.6051 |
| | $2d_{5/2}\,2d_{5/2}$ | | | 0.2371 |
| | $2d_{5/2}\,2d_{3/2}$ | | | -0.3063 |
| | $2d_{5/2}\,3s_{1/2}$ | | | 0.4812 |
| | $2d_{3/2}\,2d_{3/2}$ | | | 0.6379 |
| | $2d_{3/2}\,3s_{1/2}$ | | | 0.7044 |
| $^{116}Cd_{0.514}(2^+)$ | $1g_{7/2}\,1g_{7/2}$ | 2 | $^{114}Cd_{gs}(0^+)$ | -0.0656 |
| | $1g_{7/2}\,2d_{5/2}$ | | | 0.0090 |
| | $1g_{7/2}\,2d_{3/2}$ | | | 0.1364 |
| | $2d_{5/2}\,2d_{5/2}$ | | | -0.0318 |
| | $2d_{5/2}\,2d_{3/2}$ | | | 0.0384 |
| | $2d_{5/2}\,3s_{1/2}$ | | | -0.0603 |
| | $2d_{3/2}\,2d_{3/2}$ | | | 0.0139 |
| | $2d_{3/2}\,3s_{1/2}$ | | | -0.0216 |
| | $1g_{7/2}\,1g_{7/2}$ | 0 | $^{114}Cd_{0.558}(2^+)$ | 0.4938 |
| | $2d_{5/2}\,2d_{5/2}$ | | | 0.2080 |
| | $2d_{3/2}\,2d_{3/2}$ | | | 0.3411 |
| | $3s_{1/2}\,3s_{1/2}$ | | | 0.2514 |
| $^{116}Cd_{1.213}(2^+)$ | $1g_{7/2}\,1g_{7/2}$ | 2 | $^{114}Cd_{gs}(0^+)$ | 0.0245 |
| | $1g_{7/2}\,2d_{5/2}$ | | | -0.0548 |
| | $1g_{7/2}\,2d_{3/2}$ | | | -0.2077 |
| | $2d_{5/2}\,2d_{5/2}$ | | | -0.0242 |
| | $2d_{5/2}\,2d_{3/2}$ | | | 0.0351 |
| | $2d_{5/2}\,3s_{1/2}$ | | | -0.0373 |
| | $2d_{3/2}\,2d_{3/2}$ | | | -0.0905 |
| | $2d_{3/2}\,3s_{1/2}$ | | | -0.1120 |
| | $1g_{7/2}\,1g_{7/2}$ | 4 | $^{114}Cd_{0.558}(2^+)$ | -0.0005 |
| | $1g_{7/2}\,2d_{5/2}$ | | | 0.0917 |
| | $1g_{7/2}\,2d_{3/2}$ | | | 0.1955 |
| | $1g_{7/2}\,3s_{1/2}$ | | | 0.2414 |
| | $2d_{5/2}\,2d_{5/2}$ | | | 0.0224 |
| | $2d_{5/2}\,2d_{3/2}$ | | | -0.1389 |
| $^{116}Cd_{1.219}(4^+)$ | $1g_{7/2}\,1g_{7/2}$ | 4 | $^{114}Cd_{gs}(0^+)$ | 0.0510 |
| | $1g_{7/2}\,2d_{5/2}$ | | | 0.0316 |
| | $1g_{7/2}\,2d_{3/2}$ | | | 0.1644 |
| | $1g_{7/2}\,3s_{1/2}$ | | | 0.1110 |
| | $2d_{5/2}\,2d_{5/2}$ | | | 0.0006 |
| | $2d_{5/2}\,2d_{3/2}$ | | | -0.0281 |
| | $1g_{7/2}\,1g_{7/2}$ | 2 | $^{114}Cd_{0.558}(2^+)$ | -0.0318 |
| | $1g_{7/2}\,2d_{5/2}$ | | | 0.0060 |
| | $1g_{7/2}\,2d_{3/2}$ | | | 0.2261 |
| | $2d_{5/2}\,2d_{5/2}$ | | | -0.0162 |
| | $2d_{5/2}\,2d_{3/2}$ | | | 0.0177 |
| | $2d_{5/2}\,3s_{1/2}$ | | | -0.0311 |
| | $2d_{3/2}\,2d_{3/2}$ | | | 0.1195 |
| | $2d_{3/2}\,3s_{1/2}$ | | | 0.0512 |
| $^{116}Cd_{1.283}(0^+)$ | $1g_{7/2}\,1g_{7/2}$ | 0 | $^{114}Cd_{gs}(0^+)$ | 0.3102 |
| | $2d_{5/2}\,2d_{5/2}$ | | | -0.0173 |
| | $2d_{3/2}\,2d_{3/2}$ | | | -0.2593 |
| | $3s_{1/2}\,3s_{1/2}$ | | | -0.0635 |
| | $1g_{7/2}\,1g_{7/2}$ | 2 | $^{114}Cd_{0.558}(2^+)$ | -0.4376 |
| | $1g_{7/2}\,2d_{5/2}$ | | | -0.0086 |
| | $1g_{7/2}\,2d_{3/2}$ | | | 0.0261 |
| | $2d_{5/2}\,2d_{5/2}$ | | | 0.0309 |
| | $2d_{5/2}\,2d_{3/2}$ | | | -0.1127 |
| | $2d_{5/2}\,3s_{1/2}$ | | | 0.1080 |
| | $2d_{3/2}\,2d_{3/2}$ | | | 0.4819 |
| | $2d_{3/2}\,3s_{1/2}$ | | | 0.3839 |

**Table A4.** Two-neutron spectroscopic amplitudes for the target overlaps using the Microscopic IBM-2 method.

| Initial state | $j_1j_2$ | $J_{12}$ | Final state | S. A. |
|---|---|---|---|---|
| $^{116}Cd_{gs}(0^+)$ | $1g_{7/2}\,1g_{7/2}$ | 0 | $^{114}Cd_{gs}(0^+)$ | 0.7117 |
| | $2d_{5/2}\,2d_{5/2}$ | | | 0.7438 |
| | $2d_{3/2}\,2d_{3/2}$ | | | 0.6479 |
| | $3s_{1/2}\,3s_{1/2}$ | | | 0.5174 |
| | $1h_{11/2}\,1h_{11/2}$ | | | -1.1672 |
| | $1g_{7/2}\,1g_{7/2}$ | 2 | $^{114}Cd_{0.558}(2^+)$ | 0.1482 |
| | $2d_{5/2}\,1g_{7/2}$ | | | -0.0780 |
| | $2d_{3/2}\,1g_{7/2}$ | | | -0.1288 |
| | $2d_{5/2}\,2d_{5/2}$ | | | 0.1493 |
| | $2d_{3/2}\,2d_{5/2}$ | | | 0.0648 |
| | $3s_{1/2}\,2d_{5/2}$ | | | -0.1675 |
| | $2d_{3/2}\,2d_{3/2}$ | | | 0.1000 |
| | $3s_{1/2}\,2d_{3/2}$ | | | -0.1348 |
| | $1h_{11/2}\,1h_{11/2}$ | | | -0.1895 |
| $^{116}Cd_{0.514}(2^+)$ | $1g_{7/2}\,1g_{7/2}$ | 2 | $^{114}Cd_{gs}(0^+)$ | -0.1284 |
| | $2d_{5/2}\,1g_{7/2}$ | | | 0.0676 |
| | $2d_{3/2}\,1g_{7/2}$ | | | 0.1115 |
| | $2d_{5/2}\,2d_{5/2}$ | | | -0.1293 |
| | $2d_{3/2}\,2d_{5/2}$ | | | -0.0561 |
| | $3s_{1/2}\,2d_{5/2}$ | | | 0.1451 |
| | $2d_{3/2}\,2d_{3/2}$ | | | -0.0866 |
| | $3s_{1/2}\,2d_{3/2}$ | | | 0.1167 |
| | $1h_{11/2}\,1h_{11/2}$ | | | 0.1641 |
| | $1g_{7/2}\,1g_{7/2}$ | 0 | $^{114}Cd_{0.558}(2^+)$ | 0.2612 |
| | $2d_{5/2}\,1g_{7/2}$ | | | 0.4271 |
| | $2d_{3/2}\,1g_{7/2}$ | | | 0.9057 |
| | $2d_{5/2}\,2d_{5/2}$ | | | 0.3428 |
| | $2d_{3/2}\,2d_{5/2}$ | | | 0.9634 |
| | $3s_{1/2}\,2d_{5/2}$ | | | 0.7589 |
| | $2d_{3/2}\,2d_{3/2}$ | | | 1.0196 |
| | $3s_{1/2}\,2d_{3/2}$ | | | 1.2374 |
| | $1h_{11/2}\,1h_{11/2}$ | | | -0.9410 |

**Table A5.** Two-proton spectroscopic amplitudes for the target overlaps using the *88Sr45* interaction.

| Initial state | $j_1j_2$ | $J_{12}$ | Final state | S. A. |
|---|---|---|---|---|
| $^{116}Cd_{gs}(0^+)$ | $2p_{1/2}\,2p_{1/2}$ | 0 | $^{118}Sn_{gs}(0^+)$ | 0.38286 |
| | $1g_{9/2}\,1g_{9/2}$ | | | -0.83524 |
| | $1g_{7/2}\,1g_{7/2}$ | | | -0.21101 |
| | $2d_{5/2}\,2d_{5/2}$ | | | -0.17230 |
| | $1g_{9/2}\,1g_{9/2}$ | 2 | $^{118}Sn_{1.230}(2^+)$ | -0.13472 |
| | $1g_{9/2}\,1g_{7/2}$ | | | 0.01588 |
| | $1g_{9/2}\,2d_{5/2}$ | | | 0.10435 |
| | $1g_{7/2}\,1g_{7/2}$ | | | 0.01632 |
| | $1g_{7/2}\,2d_{5/2}$ | | | -0.00331 |
| | $2d_{5/2}2d_{5/2}$ | | | -0.00381 |
| | $2p_{1/2}\,2p_{1/2}$ | 0 | $^{118}Sn_{1.758}(0^+)$ | 0.00459 |
| | $1g_{9/2}\,1g_{9/2}$ | | | 0.01474 |
| | $1g_{7/2}\,1g_{7/2}$ | | | 0.00042 |
| | $2d_{5/2}\,2d_{5/2}$ | | | -0.00496 |
| | $1g_{9/2}\,1g_{9/2}$ | 2 | $^{118}Sn_{2.043}(2^+)$ | 0.01189 |
| | $1g_{9/2}\,1g_{7/2}$ | | | -0.00856 |
| | $1g_{9/2}\,2d_{5/2}$ | | | -0.10630 |
| | $1g_{7/2}\,1g_{7/2}$ | | | -0.01567 |
| | $1g_{7/2}\,2d_{5/2}$ | | | 0.00716 |
| | $2d_{5/2}2d_{5/2}$ | | | -0.01321 |
| | $2p_{1/2}\,2p_{1/2}$ | 0 | $^{118}Sn_{2.057}(0^+)$ | 0.00173 |
| | $1g_{9/2}\,1g_{9/2}$ | | | 0.01039 |
| | $1g_{7/2}\,1g_{7/2}$ | | | 0.01355 |
| | $2d_{5/2}\,2d_{5/2}$ | | | 0.05993 |
| | $1g_{9/2}\,1g_{9/2}$ | 4 | $^{118}Sn_{2.280}(4^+)$ | 0.01300 |
| | $1g_{9/2}\,1g_{7/2}$ | | | -0.02146 |
| | $1g_{9/2}\,2d_{5/2}$ | | | -0.09033 |
| | $1g_{7/2}\,1g_{7/2}$ | | | -0.00587 |
| | $1g_{7/2}\,2d_{5/2}$ | | | 0.00626 |
| | $2d_{5/2}2d_{5/2}$ | | | -0.01457 |
| $^{116}Cd_{0.513}(2^+)$ | $1g_{9/2}\,1g_{9/2}$ | 2 | $^{118}Sn_{gs}(0^+)$ | -0.96741 |
| | $1g_{9/2}\,1g_{7/2}$ | | | 0.00368 |
| | $1g_{9/2}\,2d_{5/2}$ | | | 0.16224 |
| | $1g_{7/2}\,1g_{7/2}$ | | | -0.00045 |
| | $1g_{7/2}\,2d_{5/2}$ | | | 0.01298 |
| | $2d_{5/2}2d_{5/2}$ | | | -0.04604 |
| | $2p_{1/2}\,2p_{1/2}$ | 0 | $^{118}Sn_{1.230}(2^+)$ | 0.33027 |
| | $1g_{9/2}\,1g_{9/2}$ | | | -0.75036 |
| | $1g_{7/2}\,1g_{7/2}$ | | | -0.19006 |
| | $2d_{5/2}\,2d_{5/2}$ | | | -0.15916 |
| | $1g_{9/2}\,1g_{9/2}$ | 2 | $^{118}Sn_{1.758}(0^+)$ | 0.03372 |
| | $1g_{9/2}\,1g_{7/2}$ | | | 0.00256 |
| | $1g_{9/2}\,2d_{5/2}$ | | | 0.02982 |
| | $1g_{7/2}\,1g_{7/2}$ | | | 0.00731 |
| | $1g_{7/2}\,2d_{5/2}$ | | | -0.00208 |
| | $2d_{5/2}2d_{5/2}$ | | | -0.00563 |
| | $1g_{9/2}\,1g_{9/2}$ | 2 | $^{118}Sn_{2.043}(2^+)$ | 0.07800 |
| | $1g_{9/2}\,1g_{7/2}$ | | | -0.01443 |
| | $1g_{9/2}\,2d_{5/2}$ | | | -0.15499 |
| | $1g_{7/2}\,1g_{7/2}$ | | | -0.01924 |
| | $1g_{7/2}\,2d_{5/2}$ | | | 0.00925 |
| | $2d_{5/2}2d_{5/2}$ | | | -0.00517 |
| | $1g_{9/2}\,1g_{9/2}$ | 2 | $^{118}Sn_{2.057}(0^+)$ | -0.00844 |
| | $1g_{9/2}\,1g_{7/2}$ | | | -0.00058 |
| | $1g_{9/2}\,2d_{5/2}$ | | | 0.00444 |
| | $1g_{7/2}\,1g_{7/2}$ | | | 0.00147 |
| | $1g_{7/2}\,2d_{5/2}$ | | | -0.00051 |
| | $2d_{5/2}2d_{5/2}$ | | | 0.05409 |
| | $1g_{9/2}\,1g_{9/2}$ | 2 | $^{118}Sn_{2.280}(4^+)$ | 0.07787 |
| | $1g_{9/2}\,1g_{7/2}$ | | | -0.02464 |
| | $1g_{9/2}\,2d_{5/2}$ | | | -0.18130 |
| | $1g_{7/2}\,1g_{7/2}$ | | | -0.02999 |
| | $1g_{7/2}$ $2d_{5/2}$ | | | 0.00982 |
| | $2d_{5/2}2d_{5/2}$ | | | 0.00013 |
| | $1g_{9/2}$ $1g_{9/2}$ | | | -1.32477 |
| | $1g_{9/2}$ $1g_{7/2}$ | | | -0.04166 |
| | $1g_{9/2}$ $2d_{5/2}$ | 2 | $^{118}Sn_{gs}(0^+)$ | -0.23671 |
| | $1g_{7/2}$ $1g_{7/2}$ | | | -0.06459 |
| | $1g_{7/2}$ $2d_{5/2}$ | | | 0.04614 |
| | $2d_{5/2}2d_{5/2}$ | | | -0.08122 |
| | $1g_{9/2}$ $1g_{9/2}$ | | | 0.54390 |
| | $1g_{9/2}$ $1g_{7/2}$ | | | 0.00164 |
| | $1g_{9/2}$ $2d_{5/2}$ | 2 | $^{118}Sn_{1.230}(2^+)$ | 0.09659 |
| | $1g_{7/2}$ $1g_{7/2}$ | | | 0.01504 |
| | $1g_{7/2}$ $2d_{5/2}$ | | | -0.01849 |
| | $2d_{5/2}2d_{5/2}$ | | | 0.03332 |
| | $1g_{9/2}$ $1g_{9/2}$ | | | 0.03111 |
| | $1g_{9/2}$ $1g_{7/2}$ | | | -0.00658 |
| | $1g_{9/2}$ $2d_{5/2}$ | 2 | $^{118}Sn_{1.758}(0^+)$ | -0.03870 |
| | $1g_{7/2}$ $1g_{7/2}$ | | | -0.00490 |
| | $1g_{7/2}$ $2d_{5/2}$ | | | 0.00603 |
| $^{116}Cd_{1.213}(2^+)$ | $2d_{5/2}2d_{5/2}$ | | | -0.00428 |
| | $1g_{9/2}$ $1g_{9/2}$ | | | -0.13610 |
| | $1g_{9/2}$ $1g_{7/2}$ | | | 0.00905 |
| | $1g_{9/2}$ $2d_{5/2}$ | 2 | $^{118}Sn_{2.043}(2^+)$ | -0.10260 |
| | $1g_{7/2}$ $1g_{7/2}$ | | | -0.00405 |
| | $1g_{7/2}$ $2d_{5/2}$ | | | 0.01095 |
| | $2d_{5/2}2d_{5/2}$ | | | -0.01829 |
| | $1g_{9/2}$ $1g_{9/2}$ | | | 0.00605 |
| | $1g_{9/2}$ $1g_{7/2}$ | | | 0.00287 |
| | $1g_{9/2}$ $2d_{5/2}$ | 2 | $^{118}Sn_{2.057}(0^+)$ | 0.02546 |
| | $1g_{7/2}$ $1g_{7/2}$ | | | -0.00201 |
| | $1g_{7/2}$ $2d_{5/2}$ | | | -0.01198 |
| | $2d_{5/2}2d_{5/2}$ | | | 0.00541 |
| | $1g_{9/2}$ $1g_{9/2}$ | | | 0.00071 |
| | $1g_{9/2}$ $1g_{7/2}$ | | | 0.00325 |
| | $1g_{9/2}$ $2d_{5/2}$ | 2 | $^{118}Sn_{2.280}(4^+)$ | 0.09836 |
| | $1g_{7/2}$ $1g_{7/2}$ | | | 0.01145 |
| | $1g_{7/2}$ $2d_{5/2}$ | | | -0.00985 |
| | $2d_{5/2}2d_{5/2}$ | | | 0.00255 |
| | $1g_{9/2}$ $1g_{9/2}$ | | | -1.95555 |
| | $1g_{9/2}$ $1g_{7/2}$ | | | -0.02777 |
| | $1g_{9/2}$ $2d_{5/2}$ | 4 | $^{118}Sn_{gs}(0^+)$ | 0.22262 |
| | $1g_{7/2}$ $1g_{7/2}$ | | | -0.03967 |
| | $1g_{7/2}$ $2d_{5/2}$ | | | 0.03459 |
| $^{116}Cd_{1.219}(4^+)$ | $2d_{5/2}2d_{5/2}$ | | | -0.07209 |
| | $1g_{9/2}$ $1g_{9/2}$ | | | -0.66457 |
| | $1g_{9/2}$ $1g_{7/2}$ | | | -0.03940 |
| | $1g_{9/2}$ $2d_{5/2}$ | 2 | $^{118}Sn_{1.230}(2^+)$ | 0.11162 |
| | $1g_{7/2}$ $1g_{7/2}$ | | | -0.01562 |
| | $1g_{7/2}$ $2d_{5/2}$ | | | 0.01137 |
| | $2d_{5/2}2d_{5/2}$ | | | -0.04840 |
| | $1g_{9/2}\ 1g_{9/2}$ | 4 | $^{118}Sn_{1.758}(0^+)$ | 0.04971 |
| | $1g_{9/2}\ 1g_{7/2}$ | | | 0.00062 |
| | $1g_{9/2}\ 2d_{5/2}$ | | | 0.03495 |
| | $1g_{7/2}\ 1g_{7/2}$ | | | 0.00154 |
| | $1g_{7/2}\ 2d_{5/2}$ | | | 0.00722 |
| | $2d_{5/2}2d_{5/2}$ | | | -0.01450 |
| | $1g_{9/2}\ 1g_{9/2}$ | 4 | $^{118}Sn_{2.043}(2^+)$ | 0.08512 |
| | $1g_{9/2}\ 1g_{7/2}$ | | | -0.01166 |
| | $1g_{9/2}\ 2d_{5/2}$ | | | -0.18736 |
| | $1g_{7/2}\ 1g_{7/2}$ | | | -0.00130 |
| | $1g_{7/2}\ 2d_{5/2}$ | | | 0.00811 |
| | $2d_{5/2}2d_{5/2}$ | | | 0.00798 |
| | $1g_{9/2}\ 1g_{9/2}$ | 4 | $^{118}Sn_{2.057}(0^+)$ | -0.01005 |
| | $1g_{9/2}\ 1g_{7/2}$ | | | -0.00161 |
| | $1g_{9/2}\ 2d_{5/2}$ | | | -0.03427 |
| | $1g_{7/2}\ 1g_{7/2}$ | | | -0.00186 |
| | $1g_{7/2}\ 2d_{5/2}$ | | | -0.00020 |
| | $2d_{5/2}2d_{5/2}$ | | | 0.03738 |
| | $2p_{1/2}\ 2p_{1/2}$ | 0 | $^{118}Sn_{2.280}(4^+)$ | -0.06383 |
| | $1g_{9/2}\ 1g_{9/2}$ | | | 0.20448 |
| | $1g_{7/2}\ 1g_{7/2}$ | | | 0.06197 |
| | $2d_{5/2}\ 2d_{5/2}$ | | | 0.09501 |
| $^{116}Cd_{1.283}(0^+)$ | $2p_{1/2}\ 2p_{1/2}$ | 0 | $^{118}Sn_{gs}(0^+)$ | 0.87657 |
| | $1g_{9/2}\ 1g_{9/2}$ | | | 0.19735 |
| | $1g_{7/2}\ 1g_{7/2}$ | | | -0.01055 |
| | $2d_{5/2}\ 2d_{5/2}$ | | | -0.01330 |
| | $1g_{9/2}\ 1g_{9/2}$ | 2 | $^{118}Sn_{1.230}(2^+)$ | 0.14828 |
| | $1g_{9/2}\ 1g_{7/2}$ | | | -0.00448 |
| | $1g_{9/2}\ 2d_{5/2}$ | | | -0.04743 |
| | $1g_{7/2}\ 1g_{7/2}$ | | | 0.00502 |
| | $1g_{7/2}\ 2d_{5/2}$ | | | -0.00420 |
| | $2d_{5/2}2d_{5/2}$ | | | 0.01191 |
| | $2p_{1/2}\ 2p_{1/2}$ | 0 | $^{118}Sn_{1.758}(0^+)$ | 0.03405 |
| | $1g_{9/2}\ 1g_{9/2}$ | | | -0.03623 |
| | $1g_{7/2}\ 1g_{7/2}$ | | | -0.01133 |
| | $2d_{5/2}\ 2d_{5/2}$ | | | -0.00015 |
| | $1g_{9/2}\ 1g_{9/2}$ | 2 | $^{118}Sn_{2.043}(2^+)$ | 0.02884 |
| | $1g_{9/2}\ 1g_{7/2}$ | | | 0.00365 |
| | $1g_{9/2}\ 2d_{5/2}$ | | | 0.05628 |
| | $1g_{7/2}\ 1g_{7/2}$ | | | 0.00257 |
| | $1g_{7/2}\ 2d_{5/2}$ | | | -0.00010 |
| | $2d_{5/2}2d_{5/2}$ | | | 0.00223 |
| | $2p_{1/2}\ 2p_{1/2}$ | 0 | $^{118}Sn_{2.057}(0^+)$ | 0.04626 |
| | $1g_{9/2}\ 1g_{9/2}$ | | | -0.03770 |
| | $1g_{7/2}\ 1g_{7/2}$ | | | -0.02336 |
| | $2d_{5/2}\ 2d_{5/2}$ | | | -0.05215 |
| | $1g_{9/2}\ 1g_{9/2}$ | 4 | $^{118}Sn_{2.280}(4^+)$ | -0.02126 |
| | $1g_{9/2}\ 1g_{7/2}$ | | | 0.00904 |
| | $1g_{9/2}\ 2d_{5/2}$ | | | 0.04142 |
| | $1g_{7/2}\ 1g_{7/2}$ | | | 0.00053 |
| | $1g_{7/2}\,2d_{5/2}$ | | | -0.00032 |
| | $2d_{5/2}2d_{5/2}$ | | | 0.00225 |

**Table A6.** Two-proton spectroscopic amplitudes for the target overlaps using the Microscopic IBM-2 method.

| **Initial state** | **$j_1j_2$** | **$J_{12}$** | **Final state** | **S. A.** |
|---|---|---|---|---|
| $^{116}Cd_{gs}(0^+)$ | $1f_{5/2}\,1f_{5/2}$ | 0 | $^{118}Sn_{gs}(0^+)$ | 0.1889 |
| | $2p_{3/2}\,2p_{3/2}$ | | | 0.1767 |
| | $2p_{1/2}\,2p_{1/2}$ | | | 0.2670 |
| | $1g_{9/2}\,1g_{9/2}$ | | | -0.7540 |
| | $1g_{7/2}\,1g_{7/2}$ | | | -0.1193 |
| | $2d_{5/2}\,2d_{5/2}$ | | | 0.0174 |
| | $1f_{5/2}\,1f_{5/2}$ | 2 | $^{118}Sn_{1.23}(2^+)$ | 0.0166 |
| | $2p_{3/2}1f_{5/2}$ | | | 0.0128 |
| | $2p_{1/2}1f_{5/2}$ | | | -0.0355 |
| | $2p_{3/2}\,2p_{3/2}$ | | | 0.0151 |
| | $2p_{1/2}2p_{3/2}$ | | | -0.0334 |
| | $1g_{9/2}\,1g_{9/2}$ | | | -0.2084 |
| | $1g_{9/2}\,1g_{7/2}$ | | | 0.0076 |
| | $1g_{9/2}\,2d_{5/2}$ | | | -0.0367 |
| | $1g_{7/2}1g_{7/2}$ | | | -0.0094 |
| | $2d_{5/2}1g_{7/2}$ | | | -0.0050 |
| | $2d_{5/2}2d_{5/2}$ | | | -0.0095 |
| | $1f_{5/2}\,1f_{5/2}$ | 0 | $^{118}Sn_{1.758}(0^+)$ | -0.0199 |
| | $2p_{3/2}\,2p_{3/2}$ | | | -0.0186 |
| | $2p_{1/2}\,2p_{1/2}$ | | | -0.0281 |
| | $1g_{9/2}\,1g_{9/2}$ | | | 0.0793 |
| | $1g_{7/2}\,1g_{7/2}$ | | | 0.0125 |
| | $2d_{5/2}\,2d_{5/2}$ | | | -0.0018 |
| | $1f_{5/2}\,1f_{5/2}$ | 2 | $^{118}Sn_{2.043}(2^+)$ | -0.0038 |
| | $2p_{3/2}1f_{5/2}$ | | | -0.0030 |
| | $2p_{1/2}1f_{5/2}$ | | | 0.0082 |
| | $2p_{3/2}\,2p_{3/2}$ | | | -0.0035 |
| | $2p_{1/2}2p_{3/2}$ | | | 0.0077 |
| | $1g_{9/2}\,1g_{9/2}$ | | | 0.0482 |
| | $1g_{9/2}\,1g_{7/2}$ | | | -0.0018 |
| | $1g_{9/2}\,2d_{5/2}$ | | | 0.0085 |
| | $1g_{7/2}1g_{7/2}$ | | | 0.0022 |
| | $2d_{5/2}1g_{7/2}$ | | | 0.0012 |
| | $2d_{5/2}2d_{5/2}$ | | | 0.0022 |
| | $1f_{5/2}\,1f_{5/2}$ | 0 | $^{118}Sn_{2.057}(0^+)$ | -0.0144 |
| | $2p_{3/2}\,2p_{3/2}$ | | | -0.0135 |
| | $2p_{1/2}\,2p_{1/2}$ | | | -0.0204 |
| | $1g_{9/2}\,1g_{9/2}$ | | | 0.0577 |
| | $1g_{7/2}\,1g_{7/2}$ | | | 0.0091 |
| | $2d_{5/2}\,2d_{5/2}$ | | | -0.0013 |
| $^{116}Cd_{0.513}(2^+)$ | $1f_{5/2}\,1f_{5/2}$ | 2 | $^{118}Sn_{gs}(0^+)$ | -0.0288 |
| | $2p_{3/2}1f_{5/2}$ | | | 0.0222 |
| | $2p_{1/2}1f_{5/2}$ | | | 0.0618 |
| | $2p_{3/2}\,2p_{3/2}$ | | | -0.0264 |
| | $2p_{1/2}2p_{3/2}$ | | | -0.0582 |
| | $1g_{9/2}\,1g_{9/2}$ | | | 0.3626 |
| | $1g_{9/2}\,1g_{7/2}$ | | | 0.0133 |
| | $1g_{9/2}\,2d_{5/2}$ | | | 0.0639 |
| | $1g_{7/2}1g_{7/2}$ | | | 0.0163 |
| | $2d_{5/2}1g_{7/2}$ | | | -0.0087 |
| | $2d_{5/2}2d_{5/2}$ | | | 0.0166 |
| | $1f_{5/2}\,1f_{5/2}$ | | | 0.0045 |
| | $2p_{3/2}1f_{5/2}$ | | | -0.0035 |
| | $2p_{1/2}1f_{5/2}$ | | | -0.0096 |
| | $2p_{3/2}\,2p_{3/2}$ | | | 0.0041 |
| | $2p_{1/2}2p_{3/2}$ | | | 0.0091 |
| | $1g_{9/2}\,1g_{9/2}$ | 2 | $^{118}Sn_{1.758}(0^+)$ | -0.0565 |
| | $1g_{9/2}\,1g_{7/2}$ | | | -0.0021 |
| | $1g_{9/2}\,2d_{5/2}$ | | | -0.0100 |
| | $1g_{7/2}1g_{7/2}$ | | | -0.0025 |
| | $2d_{5/2}1g_{7/2}$ | | | 0.0014 |
| | $2d_{5/2}2d_{5/2}$ | | | -0.0026 |
| | $1f_{5/2}\,1f_{5/2}$ | | | 0.0023 |
| | $2p_{3/2}1f_{5/2}$ | | | -0.0018 |
| | $2p_{1/2}1f_{5/2}$ | | | -0.0050 |
| | $2p_{3/2}\,2p_{3/2}$ | | | 0.0021 |
| | $2p_{1/2}2p_{3/2}$ | | | 0.0047 |
| | $1g_{9/2}\,1g_{9/2}$ | 2 | $^{118}Sn_{2.057}(0^+)$ | -0.0293 |
| | $1g_{9/2}\,1g_{7/2}$ | | | -0.0011 |
| | $1g_{9/2}\,2d_{5/2}$ | | | -0.0052 |
| | $1g_{7/2}1g_{7/2}$ | | | -0.0013 |
| | $2d_{5/2}1g_{7/2}$ | | | 0.0007 |
| | $2d_{5/2}2d_{5/2}$ | | | -0.0013 |

**Table A7.** Two proton spectroscopic amplitudes for target overlaps using the QRPA approach.

| **Initial state** | **$j_1j_2$** | **$J_{12}$** | **Final state** | **Spect. Ampl.** |
|---|---|---|---|---|
| $^{116}Cd_{g.s.}\,(0^+)$ | $1g_{9/2}\,1g_{9/2}$ | $0^+$ | $^{118}Sn_{g.s.}\,(0^+)$ | -0.9853 |
| | $2d_{5/2}\,2d_{5/2}$ | | | 0.1708 |
| | $1g_{9/2}\,1g_{9/2}$ | | | 0.1396 |
| | $1g_{7/2}\,1g_{9/2}$ | | | 0.0367 |
| | $2d_{5/2}\,1g_{9/2}$ | | | 0.1720 |
| | $1g_{7/2}\,1g_{7/2}$ | | | 0.0616 |
| | $1g_{7/2}\,2d_{5/2}$ | | | 0.0257 |
| $^{116}Cd_{g.s.}\,(0^+)$ | $2d_{3/2}\,1g_{7/2}$ | 2 | $^{118}Sn_{1.23}\,(2^+)$ | 0.0460 |
| | $2d_{5/2}\,2d_{5/2}$ | | | 0.0826 |
| | $2d_{3/2}\,2d_{5/2}$ | | | 0.0250 |
| | $3s_{1/2}\,2d_{5/2}$ | | | -0.0454 |
| | $2d_{3/2}\,2d_{3/2}$ | | | 0.0236 |
| | $2d_{3/2}$ $3s_{1/2}$ | | | -0.0228 |
| | $1h_{11/2}$ $1h_{11/2}$ | | | 0.0467 |